\documentclass[fleqn,usenatbib]{mnras}

\usepackage{newtxtext,newtxmath}

\usepackage[T1]{fontenc}

\DeclareRobustCommand{\VAN}[3]{#2}
\let\VANthebibliography\thebibliography
\def\thebibliography{\DeclareRobustCommand{\VAN}[3]{##3}\VANthebibliography}

\usepackage{graphicx}	
\usepackage{amsmath}	
\usepackage{amssymb}	
\usepackage{subfig}
\usepackage{float}

\title[X-ray polarization of 3C 273]{Constraining the X-ray radiation origin of 3C 273 in the low state by polarization}

\author[Mingjun Liu et al.]{
Mingjun Liu,$^{1,2}$
Wenda Zhang,$^{1}$\thanks{E-mail:wdzhang@nao.cas.cn}
Weimin Yuan$^{1,2}$\\
$^{1}$National Astronomical Observatories, Chinese Academy of Sciences, 20A Datun Road, Beijing 100101, People’s Republic of China\\
$^{2}$School of Astronomy and Space Sciences, University of Chinese Academy of Sciences, 19A Yuquan Road, Beijing 100049, People’s Republic of China
}

\date{Accepted XXX. Received YYY; in original form ZZZ}

\pubyear{2023}

\begin{document}
\label{firstpage}
\pagerange{\pageref{firstpage}--\pageref{lastpage}}
\maketitle

\begin{abstract}
3C 273 is one of the nearest high-luminosity quasars. Although classified as a blazar, 3C 273 also has some features in Seyferts, whose X-ray may originate from the corona. Since both jet and corona produce power-law spectra in X-ray, the spectrum cannot completely distinguish their contributions to 3C 273 in the low state. X-ray polarimetric observations provide the chance to constrain the X-ray radiation origin of 3C 273 in the low state. We perform general relativistic radiative transfer simulations with the code \textsc{MONK} to compute the X-ray polarization in 2-10 keV from the jets, sphere coronae, and slab coronae for 3C 273. We find that the radiation from the jet in 2-10 keV has a larger polarization degree than that of the corona: the polarization degree in 2-10 keV from the corona is unpolarized, while these are 4.1\%-15.8\% for the jet with a vertical or radial magnetic field and $\leq$5.0\% for the jet with toroidal magnetic field. The X-ray polarization of the corona and jet is sensitive to optical depth and geometry, and the main driver for this dependence is the number of scatterings. These results show that X-ray polarization can effectively constrain the X-ray radiation origin of 3C 273 in the low state.
\end{abstract}




\section{Introduction}

As one of the nearest high-luminosity quasars \citep[\textit{z} = 0.158;][]{1963Natur.197.1040S}, 3C 273 has been adequately investigated from radio to $\gamma$-ray. 3C 273 is classified as a blazar, while it also has some features of Seyferts. Like many blazars, 3C 273 shows strong variability in all bands \citep{2008A&A...486..411S} and has a superluminal jet \citep{1981Natur.290..365P}. \citet{1998A&A...340...35H} and \citet{1998PASJ...50..213C} first suggested that the corona/Seyfert-like component of 3C 273 exists and can be easily observed in the low state.

3C 273 gradually transitioned into a low flux state after 1996 \citep{2006AdSpR..38.1393D}. The Seyfert-like component likely dominates the X-ray, while the jet dominates the $\gamma$-ray in the \emph{RXTE}, \emph{INTEGRAL}, and \emph{Fermi} observations of 3C 273 from 2005 to 2012 \citep{2015A&A...576A.122E}. \citet{2008A&A...479..365P} applied the corona and jet models on fitting BeppoSAX data, suggesting that the power law components in 2-10 keV with the photon index around 1.68-1.72 should mainly be from the contributions of the corona rather than the jet whose photon index was found to be around 1.5 in 0.2-200 keV during 1996-2001 \citep{2004Sci...306..998G}. Estimations of the corona properties have been given by performing spectral and timing analysis. \citet{2015ApJ...812...14M} fitted the energy spectra of 3C 273 that were extracted from \emph{NuSTAR} and \emph{INTEGRAL} observations in 2011-2012, using two models: the first one is the corona model, and the second one is the corona with jet model. For the corona model, the electron temperature ($kT_{\text e}$) is $247_{-64}^{+69}$ keV, and the optical depth ($\tau$) is $0.15_{-0.04}^{+0.08}$, while these are $12\pm0.3$ keV and $2.77\pm0.06$ for the corona with jet model, respectively. Since these two models can fit the observation well, the contributions of the corona and jet to the X-ray radiation of 3C 273 in the low state cannot be completely distinguished by the spectral analysis alone. Furthermore, \citet{2017MNRAS.469.3824K} obtained the electron temperature and optical depth of 540 keV and 0.18-0.24, respectively, fitted from \emph{XMM–Newton} in 2000-2015. This X-ray emitting region in the unresolved core of 3C 273 is around 0.02-0.05 pc, located around $0.5\pm0.4$ pc from the jet apex \citep{2016A&A...590A..61C}.

Since both the jet and the corona can produce power law X-ray spectra by the synchrotron self-Compton process (SSC) and inverse Compton scattering \citep{1991ApJ...380L..51H}, respectively, it is not easy to quantitatively distinguish the contributions of the jet and corona to the X-ray radiation of 3C 273 in the low state by energy spectral analysis alone. Polarization offers us a new window to investigate the geometrical and physical properties of 3C 273. The polarization of 3C 273 in radio and ultraviolet-optical-infrared (UV-opt-IR) bands has been observed since 1962 \citep{1962AJ.....67Q.585R} and 1963 \citep{1966ZA.....64..181W}, respectively. 3C 273 has an extended large scale jet containing many knots and an unresolved core that may have the accretion disk, corona, and small scale jet \citep{1998A&ARv...9....1C}. The core has weakly polarized radiation from radio to UV band except for the mm band. Due to the high angular resolution of the radio interferometric observation, the small scale jet in the core can be spatially resolved in the radio band. The small scale jet has a higher polarization degree in the mm band, such as 4.9\%-10.4\% in 15 and 22 GHz in 1996 \citep{2002ApJ...568...99H}, 1.4\%-33.1\% in 7 mm (43 GHz) during 1998-2001 \citep{2007AJ....134..799J}, and 6.7\%-22.3\% in 43 and 86 GHz in 2002 \citep{2005ApJ...633L..85A}, while the unresolved part in the radio band in the core still has low polarized radiation \citep{2005ApJ...633L..85A, 2005AJ....130.1418J, 2005ASPC..340..183J, 2007AJ....134..799J, 2017A&A...599A..34R, 2019A&A...623A.111H, 2020A&A...636A..62K}. Therefore, the unresolved core is polarized in the mm band due to the small scale jet \citep{2016ApJ...817..131H, 2022arXiv221013819P}. The UV-opt-IR band polarization degrees of the core are hardly over 1\% since 1963 \citep{1966ZA.....64..181W, 1969ApJ...155.1113L, 1963IBVS...31....1M, 1968ApJ...151..769A, 1977ApJ...215L.107K, 1979Natur.280..215K, 1984ApJ...279..485S, 1987ApJS...64..459S, 1989ApJ...347...96I, 1992ApJ...396L..19D, 2007AJ....134..799J, 2020ApJ...897...18L}. These low polarization degrees in the UV-opt-IR band of the core were used to be interpreted as galactic interstellar polarization \citep{1966ZA.....64..181W}. However, it was later demonstrated to be intrinsic \citep{1989ApJ...347...96I}, which may originate from scattering by dust in the torus or free electrons in the core \citep{1984ApJ...279..485S, 2007A&A...465..129G, 2018A&A...614A.120S}. The large scale jet has higher polarized radiation compared with the core. The radio band polarization degrees of large scale jet are 10\%-20\% in 408-4835 MHz in 1984-1985 \citep{1993A&A...267..347C}. The optical polarization degrees of the large scale jet are $3.7\pm4.1\%$ in 1977 \citep{1978ApJ...220L..31S}, $7.5\pm2.2\%$ in 1983-1984 \citep{1986A&A...154...15R}, $15.1\pm3.9\%$ in 1987 \citep{1987MNRAS.228P..35S}, and $\lesssim$20\% in 1985-1988 \citep{1991A&A...252..458R, 1996A&A...314..414R}. These are used to be interpreted as Thomson scattering \citep{1992MNRAS.254..488M} or inverse Compton scattering of background radiation \citep{1978ApJ...220L..31S}. The consistency between radio and optical polarization \citep{2006ApJ...648..910U} and some observations \citep{1987MNRAS.228P..35S, 2014ApJ...780L..27M} tend to support the synchrotron origin of the polarization in the large scale jet. However, due to the lack of X-ray polarimeters, the X-ray polarization of 3C 273 is rarely reported. 

The newly launched X-ray polarimeter satellite, the Imaging X-ray Polarimetry Explorer \citep[\emph{IXPE},][]{2016SPIE.9905E..17W}, working in the 2-8 keV bands, makes high-sensitivity polarization investigations in the soft X-ray band possible. The enhanced X-ray Timing and Polarimetry mission \citep[\emph{eXTP},][]{2016SPIE.9905E..1QZ} to be launched in 2027, also has a polarimetry focusing array working in 2-8 keV \citep{2019SCPMA..6229502Z}. Many active galactic nuclei (AGN) are the targets of these two missions \citep{2016SPIE.9905E..17W, 2016SPIE.9905E..1QZ, 2017sf2a.conf..173M, 2022arXiv220709338M}. In 2022, the 95.28 ks observation on 3C 273 by \emph{IXPE} gives an 99 \% confidence level upper limit of 9\% on the polarization degree \citep{2023arXiv231011510M}. Since the polarization characteristics from synchrotron radiation and inverse Compton scattering are different, the X-ray polarization of 3C 273 is expected to put tighter constraints on the contributions of its jet and corona. Moreover, some X-ray polarization simulation studies on X-ray binaries \citep{2022MNRAS.514.2561G, 2022MNRAS.tmp.1852Z} and AGNs \citep{2022MNRAS.510.3674U} show that X-ray polarization can constrain the geometry of the corona.

In this work, we perform simulations to compute the X-ray polarization of the corona and jet utilizing the general relativistic Monte Carlo radiative transfer code \textsc{MONK} \citep{2019ApJ...875..148Z}, which can be used to put constraints on the dominant X-ray compositions of 3C 273 in the low state with X-ray polarimetric measurements. The methods and set up of our simulations are described in Section ~\ref{sec:method}. In Section ~\ref{sec:result}, we present the results of polarization in simulations of the corona and jet and investigate the influence of the number of scatterings on polarization. In Section ~\ref{sec:dis}, we mainly discuss the feasibility of distinguishing dominant X-ray components for 3C 273 by polarization. The main conclusions are summarized in Section ~\ref{sec:con}.

\section{Methods and parameters}
\label{sec:method}

\subsection{Methods}

\textsc{MONK} performs Monte Carlo general relativistic radiative transfer simulation including a few radiation mechanisms, such as Compton scattering, synchrotron radiation, synchrotron self-adsorption,  bremsstrahlung, and bremsstrahlung self-adsorption. The detailed process of radiative transfer can be found in \citet{2019ApJ...875..148Z}. Here we give a brief description. First, the seed photons are produced in the accretion disc or the relativistic jet according to the emissivity of the assumed emission process. We use the “superphoton” scheme, in which a superphoton is a package of several identical photons. Each superphoton is assigned a statistical weight $w$ that has the physical meaning of the number of photons emitted per unit time. Next, the wave vectors of photons are propagated along the null geodesics in the Kerr space-time carrying their orthogonal polarization vectors and the conserved Walker-Penrose constant $\kappa_{\rm wp}$ along a tiny step. In each step, we calculate the Compton scattering possibility based on Section 2.6.1 in \citet{2019ApJ...875..148Z} assuming the Klein-Nishina cross section. If scattering happens, we calculate the polarization vector before scattering by $\kappa_{\rm wp}$ in the Boyer-Lindquist frame, then transform it into the electron rest frame. After scattering, the photon energy and wave vector are re-sampled. After resampling the wave vector, we calculate Stokes parameters in the electron rest frame to get the polarization vector, then transform the wave vector and polarization vector from the electron rest frame to the Boyer-Lindquist frame to re-evaluate $\kappa_{\rm wp}$. Finally, except for the photons that are absorbed, other photons arrive at infinity, enter the black hole event horizon, or hit the accretion disc. We compute the energy and polarization spectrum using the information of the photons that arrive at infinity. We define the orientation of the 0$^{\circ}$ polarization angle as the direction along the jet or the symmetry axis of the black hole. 

To evaluate the uncertainties of the polarization degree and angle in our simulations, for each case, we perform several identical simulations, compute the polarization and angle for every simulation, and estimate the standard error:
\begin{equation}
   e_\mathrm{r} = \sqrt{\sum_{i=1}^{n}\frac{(X_i-\Bar{X})^2}{n\left(n-1\right)}},
\end{equation}
where $X_i$ is the polarization degree or polarization angle of a particular energy bin from the $i$-th simulation, $n$ is the total number of identical simulations for one case, and $\Bar{X}$ is the mean value of $X_i$. As a sanity check, we examine how $e_\mathrm{r}$ changes with the number of superphotons used in one simulation $N$. We find the error of polarization degree to be roughly proportional to $1/\sqrt{N}$, as expected for the Monte-Carlo simulation. We also notice that, in a few cases, the uncertainty of polarization angle does not decrease with $N$ but rather stays constant. This is what we expect if the radiation is intrinsically unpolarized. In practice, for each case we fit the uncertainly of the polarization angle vs. $N$ with a power-law function: $e_\mathrm{r}\propto N^{-\alpha}$, and take the cases in which $\alpha$< 0.35 as unpolarized.

We set the number of superphotons in such a way that the uncertainties of polarization degree and polarization angle are smaller than the errors of the observations of AGNs made by \emph{IXPE}. For AGNs, the smallest errors of polarization degree and polarization angle measured by \emph{IXPE} are 1.1\% \citep[IC 4329A,][]{2023MNRAS.525.5437I} and 4$^\circ$ \citep[Mrk 421,][]{2022ApJ...938L...7D}, respectively. Therefore, we set the number of superphotons to achieve the maximum standard errors of the polarization degree and angle to be 1.1\% and 4$^\circ$, respectively.

\subsection{Set up}
\label{sec:set}

In this simulation, the black hole mass of 3C 273 is taken as $4.1\times10^8$ solar masses referring to the measurement by \citet{2019ApJ...876...49Z}. The spectra of 3C 273 indicate a high spin black hole with the dimensionless spin parameter \emph{a} over 0.9 \citep{2014xru..confE..64D, 2009MNRAS.400.1521J}. Therefore, we set \emph{a} = 0.998 in the corona simulations. To investigate whether X-ray polarization can constrain spin, we also perform simulations assuming \emph{a} = 0 for the corona simulations. On the other hand, due to the large scale of the jet of 3C 273, the influence of black hole spin on radiation from the jet should be weak. Therefore, we only take \emph{a} = 0.998 for the jet simulations. 

\begin{figure*}
	\centering
	\subfloat[Sphere corona]{\label{fig:geo_a}\includegraphics[width=1.35in]{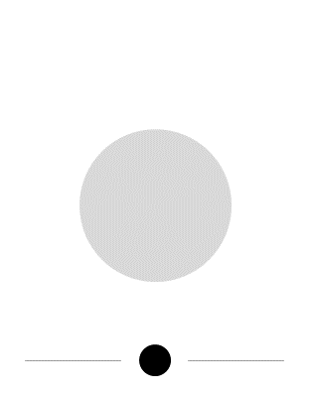}}
    \subfloat[Slab corona]{\label{fig:geo_b}\includegraphics[width=1.35in]{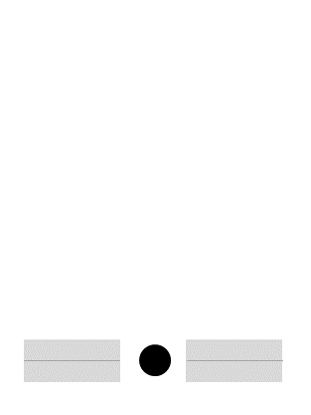}}
	\subfloat[Jet in vertical field]{\label{fig:geo_c}\includegraphics[width=1.35in]{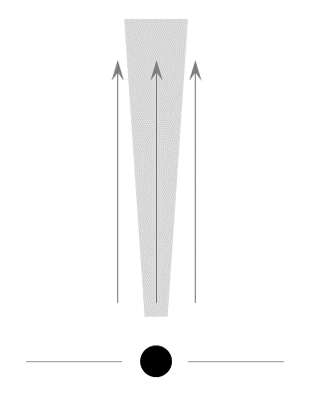}}
    \subfloat[Jet in radial field]{\label{fig:geo_d}\includegraphics[width=1.35in]{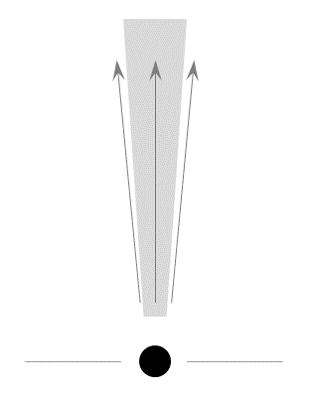}}
	\subfloat[Jet in toroidal field]{\label{fig:geo_e}\includegraphics[width=1.35in]{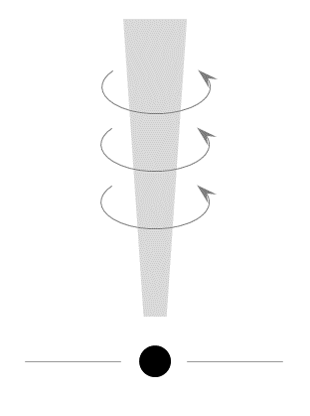}}
    \caption{The geometries of the corona and jet. The black solid circles represent the black holes. The horizontal grey lines represent the accretion disc. The grey shadows represent the corona or jet. The grey arrows represent the direction of magnetic field.}
    \label{fig:geo}
\end{figure*}

We simulate the sphere corona, slab corona, and jet in our investigation. The schematic plots of the corona and jet are shown in Fig.~\ref{fig:geo}. We choose the parameters in our simulations by fitting the observed energy spectra. More specifically, after fixing a few parameters (e.g. the black hole spin and the height of the sphere corona), we set the values of remaining parameters in such a way that the flux and photon index of the energy spectra extracted from our simulations are consistent with the observations. To compute the observed photon index and flux, we take the mean values of the photon index and flux in 2.5-10 keV of 20 \emph{XMM–Newton} observations in 2000-2015 from \citet{2017MNRAS.469.3824K} and compute the mean values for comparison. The mean values of the photon index and flux density in 2.5-10 keV are 1.59 and $8.8\times10^{-11}$ erg cm$^{-2}$ s$^{-1}$, respectively. To compare the flux density from the observations with the specific luminosity in our simulations, a luminosity distance to the source is also required. We take the cosmological parameters $\emph{H}_0=69.32$ km s$^{-1}$ Mpc$^{-1}$, $\Omega_{\rm M}=0.29$ and $\Omega_{\rm \Lambda}=0.71$, following \citet{2013ApJS..208...19H}. All measurements of the inclination angle by analyzing the superluminal motion of the jet support a face-on geometry: 0$^{\circ}$-11.9$^{\circ}$ \citep{1981Natur.290..365P}, 3.8$^{\circ}$-7.2$^{\circ}$ \citep{2016ApJ...818..195M} and 6.4$^{\circ}\pm$2.4$^{\circ}$ \citep{2017ApJ...846...98J}. However, comparing the linear polarization of the large scale jet (18\%, magnetic field parallel to the jet axis) and its shocked gas (23\%, magnetic field transverse), \citet{1994A&A...284..724C} suggested a likely larger inclination over 30$^{\circ}$. Therefore, we mainly take the inclination angle \emph{i} of 0$^{\circ}$-12$^{\circ}$ in our analysis, but also investigate other inclinations.

\begin{table}
	\centering
	\caption{The geometrical and physical parameters of the corona.}
	\label{tab:corona}
	\resizebox{\columnwidth}{!}{
	\begin{tabular}{ccccccccc}
		\hline
		 & \multicolumn{4}{c}{Radius of sphere corona (\emph{R}$_{\rm G}$)} & \multicolumn{4}{c}{Optical depth ($\tau$)}\\
		\hline
		\emph{a} & $\emph{h}=3$ & $\emph{h}=5$ & $\emph{h}=10$ & $\emph{h}=20$ & $\emph{kT}_{\rm e}$ (keV) & 50 & 100 & 200\\
		\hline
		0.998 & 1.7 & 3.7 & 6 & 10 & Sphere & 2.65 & 1.55 & 0.75\\
		0 & 0.8 & 2.8 & 7.8 & 12 & Slab & 1 & 0.5 & 0.2\\
		\hline
	\end{tabular}}
\end{table}

\begin{table}
    \caption{The geometrical and physical parameters of the jet.}
    \label{tab:jet}
    \resizebox{\linewidth}{!}{
    \begin{tabular}{cccccc}
        \hline
        \emph{a} & $\emph{N}_{\rm e}^{\rm a}$ (cm$^{-3}$) & Length of jet apex (\emph{R}$_{\rm G}$) & $p^{\rm b}$ & $\tau^{\rm c}$\\
        \hline
        0.998 & 10 & 1$\times$10$^{5}$ & 2.2 & 4.02$\times$10$^{-5}$\\       
        0.998 & 50 & 1$\times$10$^{5}$ & 2.7 & 2.01$\times$10$^{-4}$\\
        0.998 & 100 & 4$\times$10$^{4}$ & 2.2 & 1.61$\times$10$^{-4}$\\
         \hline
    \end{tabular}}
    \footnotesize{$^{\rm a}$ The number density of electrons.}\\
    \footnotesize{$^{\rm b}$ The index of the power-law electron velocities distributions.}\\
    \footnotesize{$^{\rm c}$ The optical depth.}
\end{table}

\begin{table}
    \centering
    \caption{The photon index and flux of each parameter group for the corona and jet, where \emph{h} in unit of \emph{R}$_{\rm G}$, $\emph{kT}_{\rm e}$ in unit of keV and $\emph{N}_{\rm e}$ in unit of cm$^{-3}$.}
	\label{tab:fit}
	\resizebox{\columnwidth}{!}{
    \begin{tabular}{ccccccc}
        \hline
        Corona & \multicolumn{3}{c}{Photon index in 2.5-10 keV} & \multicolumn{3}{c}{Flux in 2.5-10 keV (erg cm$^{-2}$ s$^{-1}$)}\\
        \hline
        \emph{a} = 0 & $\emph{kT}_{\rm e}=50$ & $\emph{kT}_{\rm e}=100$ & $\emph{kT}_{\rm e}=200$ & $\emph{kT}_{\rm e}=50$ & $\emph{kT}_{\rm e}=100$ & $\emph{kT}_{\rm e}=200$\\
        \hline
        $\emph{h}=3$ & 1.61 & 1.65 & 1.59 &	4.43$\times$10$^{-13}$ & 4.89$\times$10$^{-13}$ & 5.31$\times$10$^{-13}$\\
        $\emph{h}=5$ & 1.65 & 1.59 & 1.59 &	6.40$\times$10$^{-12}$ & 8.54$\times$10$^{-12}$ & 7.94$\times$10$^{-12}$\\
        $\emph{h}=10$ & 1.61 & 1.60 & 1.62 & 4.50$\times$10$^{-11}$ & 5.75$\times$10$^{-11}$ & 5.74$\times$10$^{-11}$\\
        $\emph{h}=20$ & 1.60 & 1.56 & 1.61 & 6.88$\times$10$^{-11}$ & 9.18$\times$10$^{-11}$ & 9.09$\times$10$^{-11}$\\
        Slab & 1.64 & 1.63 & 1.58 &	1.78$\times$10$^{-9}$ & 1.54$\times$10$^{-9}$ & 1.04$\times$10$^{-9}$\\
        \hline
        \emph{a} = 0.998 & $\emph{kT}_{\rm e}=50$ & $\emph{kT}_{\rm e}=100$ & $\emph{kT}_{\rm e}=200$ & $\emph{kT}_{\rm e}=50$ & $\emph{kT}_{\rm e}=100$ & $\emph{kT}_{\rm e}=200$\\
        \hline
        $\emph{h}=3$ & 1.57 & 1.58 & 1.52 & 2.30$\times$10$^{-11}$ & 2.67$\times$10$^{-11}$ & 2.90$\times$10$^{-11}$\\
        $\emph{h}=5$ & 1.52 & 1.52 & 1.56 &	6.94$\times$10$^{-11}$ & 8.87$\times$10$^{-11}$ & 8.65$\times$10$^{-11}$\\
        $\emph{h}=10$ & 1.55 & 1.54 & 1.59 & 7.08$\times$10$^{-11}$ & 8.64$\times$10$^{-11}$ & 9.02$\times$10$^{-11}$\\
        $\emph{h}=20$ & 1.49 & 1.50 & 1.63 & 7.59$\times$10$^{-11}$ & 9.71$\times$10$^{-11}$ & 8.18$\times$10$^{-11}$\\
        Slab & 1.58 & 1.57 & 1.52 &	1.63$\times$10$^{-9}$ & 1.47$\times$10$^{-9}$ & 1.09$\times$10$^{-9}$\\
        \hline
        Jet & $\emph{N}_{\rm e}=10$ & $\emph{N}_{\rm e}=50$ & $\emph{N}_{\rm e}=100$ & $\emph{N}_{\rm e}=10$ & $\emph{N}_{\rm e}=50$ & $\emph{N}_{\rm e}=100$\\
        \hline
        $\emph{B}_{\rm v}$ & 1.55 & 1.55 & 1.50 & 9.11$\times$10$^{-11}$ & 8.66$\times$10$^{-11}$ & 1.41$\times$10$^{-10}$\\
        $\emph{B}_{\rm r}$ & 1.52 & 1.55 & 1.50 &	1.17$\times$10$^{-10}$ & 8.23$\times$10$^{-11}$ & 1.35$\times$10$^{-10}$\\
        $\emph{B}_{\rm t}$ & 1.54 & 1.51 & 1.50 & 1.18$\times$10$^{-10}$ & 8.13$\times$10$^{-11}$ & 1.55$\times$10$^{-10}$\\
        \hline
    \end{tabular}}
\end{table}

\begin{itemize}
    \item For the corona simulations, the origin of the X-ray is the inverse Comptonization in the corona. Therefore, for the opacities, only Compton scattering is included. The seed photons from the accretion disc are assumed to be unpolarized and isotropic in the inertial frame of the disc. The emissivity of the accretion disc follows \citet{1973blho.conf..343N}. The disc extends from the innermost stable circular orbit (ISCO) to 100 \emph{R}$_{\rm G}$ ($\emph{R}_{\rm G}=GM/c^2$). The Eddington mass accretion rates $\dot{m}$ are taken as 0.8 and 0.2 for \emph{a} = 0 or 0.998, respectively, based on the estimations by \citet{2009MNRAS.400.1521J}. The density distribution of the corona is uniform. We assume the electrons in the corona to be thermal and obey the Maxwell-Jüttner distribution, set their thermal velocities as isotropic, and take their temperatures ($\emph{kT}_{\rm e}$) as 50, 100, and 200 keV. The weight of superphotons for corona simulation follows \citet{2019ApJ...875..148Z}.
    
    We simulate the sphere and slab coronae in our investigation. Fig.~\ref{fig:geo}(a) and (b) show schematic plots of the sphere and slab coronae, respectively. The sphere corona is above the equatorial plane and is characterized by its radius $\emph{R}_{\rm c}$ and height \emph{h} above the equatorial plane. We choose four different values of height: 3, 5, 10, and 20 $\emph{R}_{\rm G}$, and for each value of height, we choose the radius and optical depth of the corona in such a way that the low-state energy spectrum of 3C 273 (i.e. the photon index of 1.59 and flux of $8.8\times10^{-11}$ erg cm$^{-2}$ s$^{-1}$ in 2-10 keV) can be reproduced. The values of the radius and optical depth we choose for sphere coronae are shown in Tab.~\ref{tab:corona}. Since the corona cannot enter the event horizon, the size of the corona is limited by the height. Only $h=20$ $\emph{R}_{\rm G}$ of the Schwarzschild black hole and $h = 5, 10, 20$ $\emph{R}_{\rm G}$ of the Kerr black hole for the sphere corona can reach the observation mean flux, as shown in Tab.~\ref{tab:fit}.

    The slab corona fully covers and co-rotates with the accretion disc. The inner radii of the slab coronae also extend to ISCO, while their thicknesses are set as 1 \emph{R}$_{\rm G}$, following \citet{2019ApJ...875..148Z}. The values of optical depths we choose for the slab coronae for each electron temperature in Tab.~\ref{tab:corona}. However, the fluxes of the slab corona in all parameters are one order of magnitude higher than the observed flux, as shown in Tab.~\ref{tab:fit}, indicating that the slab corona should partially cover the disc if its radiation dominates the X-ray of 3C 273 in the low state.
\end{itemize}

\begin{itemize}
    \item For the jet simulations, we ignore the seed photons from the disc for the large volume and distance of the jet to the black hole in 3C 273 and only consider the X-ray from the SSC process in the jet. Therefore, Compton scattering, synchrotron radiation, and synchrotron self-adsorption are included. The jet of 3C 273 has a bulk velocity along the radial direction with the Lorentz factor $\Gamma_{\rm bulk}$ set as 10.6 \citep{2005AJ....130.1418J}. In the local rest frame of the jet, the electron velocities are assumed to be isotropic and follow the power-law distribution described by the index and the maximum and minimum Lorentz factor of this electron isotropic velocities \citep{2011ApJ...737...21L}:
    \begin{equation}
        \frac{\text dN_{\text e}}{\text d\gamma\text d\Omega_p}=\frac{N_{\text e}\left(p-1\right)}{4\pi(\gamma_{\text {min}}^{1-p}-\gamma_{\text{max}}^{1-p})}\gamma^{-p},
    \end{equation}
    where $N_{\text e}$ is the number density of electrons, $\gamma$ is the Lorentz factor of isotropic electron velocities in the local rest frame of the jet, $\Omega_p$ is the solid angle in the momentum space of electron, $p$ is the index of the power-law electron velocities distributions, $\gamma_{\text{min}}$, and $\gamma_{\text{max}}$ are the minimum and maximum values of Lorentz factor $\gamma$. The $\gamma_{\text{max}}$ is around 2.2-2.8$\times$10$^{3}$ constrained from observations \citep{2015A&A...576A.122E}. For simplicity, we take $\gamma_{\rm max}$ as 2$\times$10$^{3}$. The $\gamma_{\text{min}}$ is set as 1, which has little influence on our results. The emission coefficient of synchrotron radiation is \citep{2011ApJ...737...21L}:
    \begin{equation}
    \begin{split}
        j_{\nu}=N_e\left(\frac{e^2\nu_{\text c}}{c}\right)&\frac{3^{p/2}\left(p-1\right)\sin{\theta}}{2(p+1)(\gamma_{\text {min}}^{1-p}-\gamma_{\text{max}}^{1-p})}\Gamma\left(\frac{3p-1}{12}\right)\\
        &\Gamma\left(\frac{3p+19}{12}\right)\left(\frac{\nu}{\nu_{\text c}\sin{\theta}}\right)^{-(p-1)/2},
    \end{split}
    \end{equation}
    where $\nu$ is the frequency of photons, $e$ is the charge of electron, $\nu_{\text c}$ is the electron cyclotron frequency, $c$ is the speed of light, and $\theta$ is the pitch angle between the photon emission direction and magnetic field. The polarization degree for seed photons of the synchrotron radiation in the jet is $(p+1)/(p+7/3)$ \citep{1979rpa..book.....R}. The polarization orientations of the seed photons are orthogonal to the magnetic field. The weight of one superphotons originating from a volume element in the jet simulations can be obtained by
    \begin{equation}
        w=\frac{\sqrt{-g}\text dV}{N_\mathrm{s}}\iint \frac{j_\nu}{h_\mathrm{P}\nu}\text d\Omega \text d\nu,
    \end{equation}
    where $g$ is the determinant of the metric tensor, $\text dV$ is the size of the volume element (and in the Boyer-Lindquist coordinates $\text dV \equiv \text dr\text d\theta\text d\phi$), $N_\mathrm{s}$ is the number of superphotons generated in the volumen element, $h_\mathrm{P}$ is the Planck constant, and $\Omega$ is the solid angle in the momentum space of photon.
    
    We assume three different configurations of the magnetic field: vertical, radial, and toroidal fields \citep{2021ApJ...912...35N, 2021ApJ...914..131Y}, as shown in Figure~\ref{fig:geo}(c)-(e). Various measurements of the magnetic field strength in the jet apex of 3C 273 indicate a magnetic field strength of the order of 1G: Faraday rotation \citep{2019A&A...623A.111H}, equipartition \citep{2017MNRAS.468.4478L}, fitting of radio \citep{2008ASPC..386..451S} and X-ray spectra \citep{2015A&A...576A.122E}. Therefore, we take the magnetic field strength as 1 G. For the jet geometry, the shape of the jet is set as a truncated cone. The opening angle of the jet is taken as 2.8$^{\circ}$ \citep{2005AJ....130.1418J}. The minimum distance from the center of the black hole to the jet is set as 10 \emph{R}$_{\rm G}$, which hardly has influences on the flux. 
    
    Similarly to the corona cases, the values of a few parameters of the jet are obtained by fitting the energy spectra. We notice a spectral degeneracy for SSC radiation that the hard electron energy spectrum with low optical depth can produce the same SSC radiation as the soft one with large optical depth. The reason is that the spectral index of the SSC process is dominated by the electron energy spectrum index for the first scattering and dominated by Compton scattering for multiple scatterings. For Compton scattering, the photon spectrum hardens with increasing the number of scatterings, which means the hard electron energy spectrum with low optical depth can produce the same SSC radiation as the soft one with large optical depth under proper geometry. To check the polarization differences for this spectrum degeneracy, we set two parameter groups with relatively low optical depth ($4.02\times10^{-5}$ and $1.61\times10^{-4}$) and the same index $p$ at 2.2, and a third group with relatively high optical depth ($2.01\times10^{-4}$) and a higher index $\gamma$ at 2.7, as shown in Tab.~\ref{tab:jet}. 
    
    In our fitting, the influences of magnetic field configuration on index and flux are weak, as shown in Tab.~\ref{tab:fit}. Therefore, we take the same values of these geometrical and physical parameters for three magnetic field configurations. Beside the inclination of 0$^{\circ}$-12$^{\circ}$, we also fit the energy spectra assuming an observer’s inclination of 30$^{\circ}$-40$^{\circ}$ based on the parameters mentioned above and even change $\Gamma_{\rm bulk}$ according to \citet{1994A&A...284..724C}. However, if the whole flux originates from the jet, due to the strong relativistic beaming and small open angle, none of these parameters can produce the observed flux and photon index at the same time. If the flux in one parameter group can match the observed flux, the photon index is smaller than that of observation. If we take the photon index as the observation value, the flux is smaller than that observed. Hence, we only discuss the results from 0$^{\circ}$-12$^{\circ}$ inclination for the jet. Since we only focus on the unresolved core rather than the large scale jet on several parsecs \citep{2006ApJ...648..910U}, the jet in this work points to the jet apex in the unresolved core of 3C 273 around 1 pc \citep{2016A&A...590A..61C}.
\end{itemize}

\section{Results}
\label{sec:result}

\subsection{Corona}

\begin{figure*}
	\includegraphics[width=\textwidth]{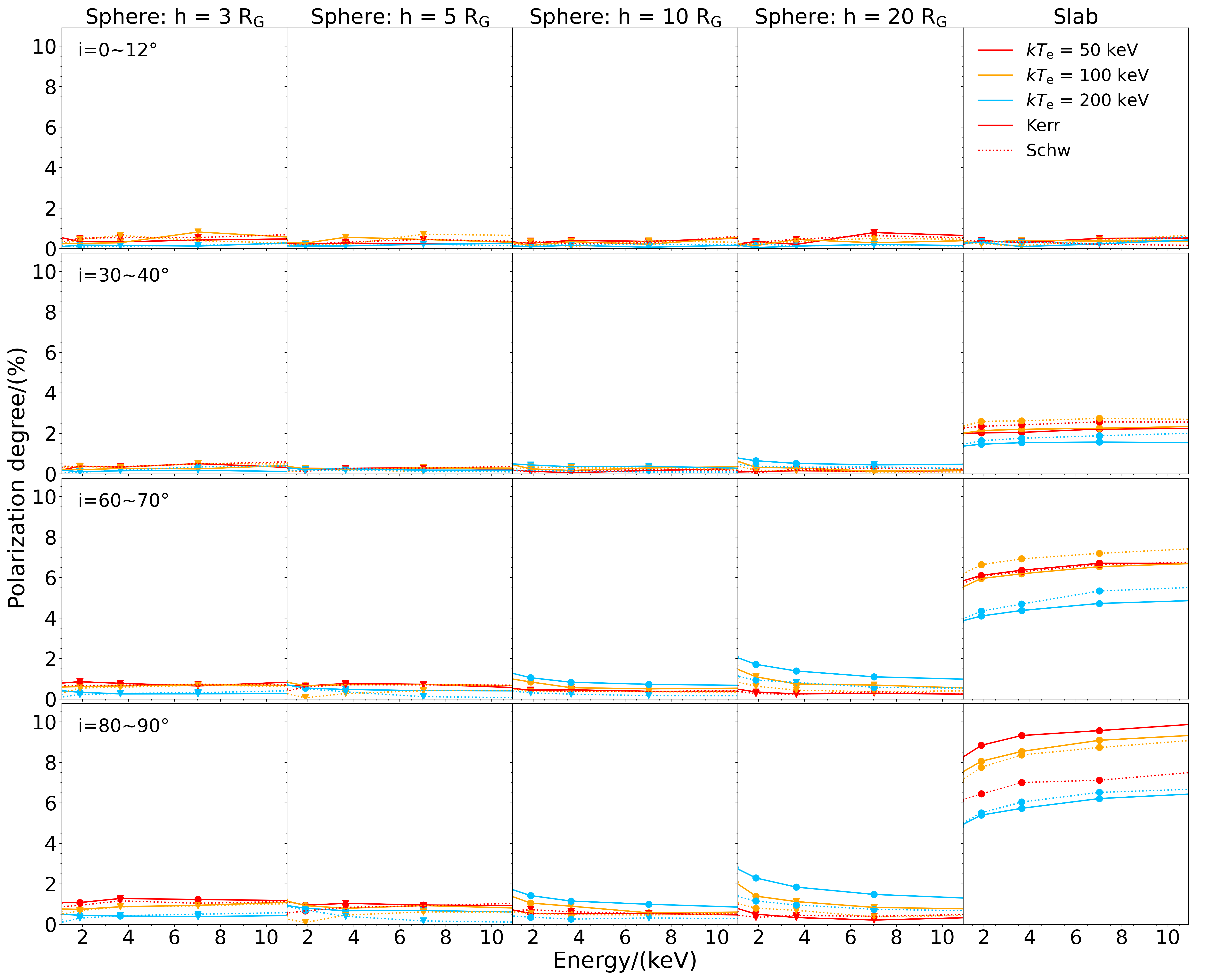}
    \caption{The 2-10 keV polarization degree of the corona. The Kerr situations are shown as dashed line. The Schwarzshild situations are shown as solid line. The electron temperature of these situations are marked by color: red, yellow, and blue for 50, 100, and 200 keV, respectively. The values of polarization degree are marked by solid circle. Some energy bins in different cases do not show polarization from the convergence of standard error with superphoton number. We use the sum of mean value of polarization degree and standard error from repeated simulations to give upper limits for these unpolarized cases, marked by solid triangle. All standard errors of polarization degree in this band for corona are below 0.24\%.}
    \label{fig:poldeg_corona}
\end{figure*}

The X-ray polarization degrees in 2-10 keV for the sphere and slab coronae at the inclinations of 0$^{\circ}$-12$^{\circ}$, 30$^{\circ}$-40$^{\circ}$, 60$^{\circ}$-70$^{\circ}$, and 80$^{\circ}$-90$^{\circ}$ are shown in Fig.~\ref{fig:poldeg_corona}. The X-rays in 2-10 keV from both sphere and slab coronae at the inclination of 0$^{\circ}$-12$^{\circ}$ are unpolarized. The X-ray radiations from sphere corona in 2-10 keV at other inclinations are also weakly polarized or unpolarized: for nearly all four parameter groups, their polarization degrees are less than $\sim$2\% and increase with the inclination. For the slab corona, the polarization degrees are larger than those of the sphere corona for the 30$^{\circ}$-90$^{\circ}$ inclinations. The polarization degree for slab corona increases with the inclination and photon energy, while it decreases with the electron temperature. For example, in the case with 50 keV electron temperature in Kerr situation at 7.03 keV, the polarization degree increases from 2.2\% for 30$^{\circ}$-40$^{\circ}$ to 9.6\% for 80$^{\circ}$-90$^{\circ}$. The effects of inclination on polarization degree mainly originate from symmetry \citep{1985A&A...143..374S}. For both the sphere and the slab coronae, the polarization degrees of the Kerr and Schwarzshild situations have little difference in the 0$^{\circ}$-12$^{\circ}$ inclination bin. Spin influences the polarization degree of the corona mainly by the size of the corona \citep{2022MNRAS.510.3674U}. As for 3C 273, the differences in corona radii between the Schwarzshild and Kerr situations are small from fitting observations. Therefore, the spin of black hole has insignificant effects on the X-ray polarization degrees of the corona for 3C 273.

\begin{figure*}
	\includegraphics[width=\textwidth]{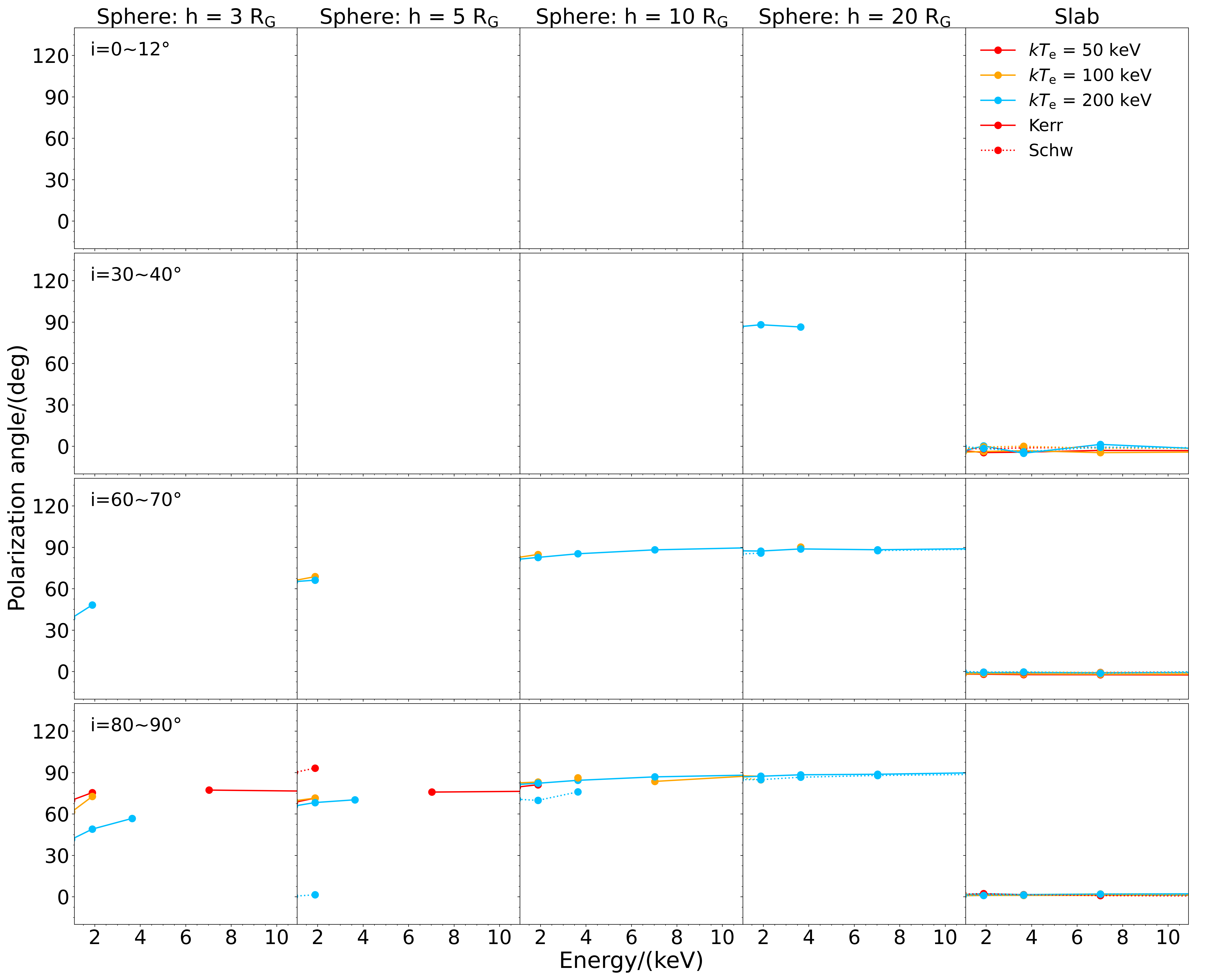}
    \caption{The 2-10 keV polarization angle of the corona. The Kerr situations are shown as dashed line. The Schwarzshild situations are shown as solid line. The electron temperature of these situations are marked by color: red, yellow, and blue for 50, 100, and 200 keV, respectively. The standard errors of polarization angle for all the energy bins with clear polarization degree in this band of corona are below 2.4$^{\circ}$.}
    \label{fig:polang_corona}
\end{figure*}

The polarization angles in 2-10 keV for the corona are shown in Fig.~\ref{fig:polang_corona}. From the large viewing inclination, the orientations of polarization are perpendicular to the black hole axis ($\sim90^{\circ}$) for the sphere coronae except the Schwarzshild case with $h = 5$ $\emph{R}_{\rm G}$ and 200 keV. For the slab coronae, the orientations of polarization are always parallel to the black hole axis ($\sim0^{\circ}$).

\begin{figure*}
  \includegraphics[width=\textwidth]{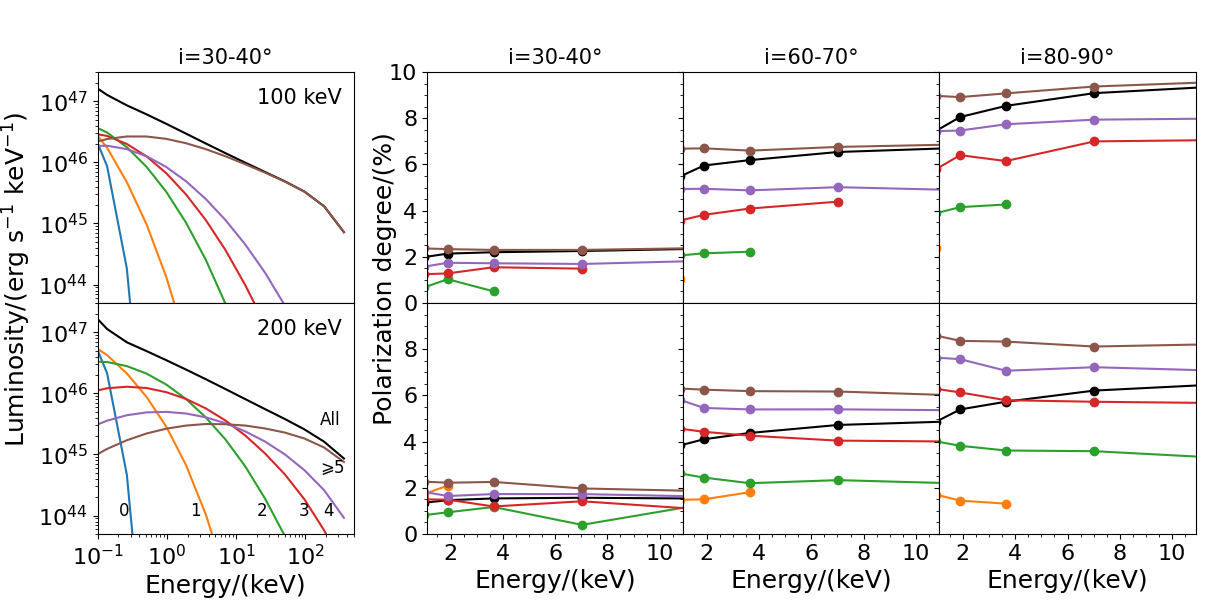}
  \caption{The flux and polarization degree of the slab corona for photons with different number of scatterings. The upper panels correspond to 100 keV coronal temperature and the lower panel corresponds to 200 keV coronal temperature. The photons that are scattered 0, 1, 2, 3, 4, and 5 or more times are represented by blue, yellow, green, red, purple, and brown lines, respectively, while the results of total emission are represented by black lines.}
  \label{fig:polsca_corona}
\end{figure*}

As mentioned above, the electron temperature can influence the polarization degree of the slab corona at large inclinations. To investigate these effects, we calculate the flux and polarization degrees from different numbers of scatterings of 100 and 200 keV for the Kerr situation. As seen in Fig.~\ref{fig:polsca_corona}, for the inverse Compton scattering in the slab corona, the polarization degree increases with the number of scatterings. For Compton scattering, the polarization degree after scattering is sensitive to the direction dispersion of incident radiations. More isotropic incident radiations with more single directions lead to a lower polarization degree of the scattered radiation. For example, the sphere corona with a larger radius and height has a higher polarization degree because it is mainly illuminated from the bottom \citep{2022MNRAS.510.3674U}, and the corona with relativistic velocity has a higher polarization degree due to the beaming of the directions of the seed photons \citep{2022MNRAS.tmp.1852Z}. For the slab coronae in our simulations, their thicknesses are much smaller than their radial extent, and their optical depths are less than 1. Therefore, for multi-scattering, the directions of incident photons from the last scattering are mainly parallel to the disc, leading to an increase in the polarization degree. For the 200 keV situation, due to the high energy gain in single scattering, to match the observed photon index, an optical depth smaller than that of the 100 keV situation is required. As a result, the 2-8 keV band is dominated by photons with a smaller number of scatterings. In 2-10 keV, the photons that are scattered at least 5 times dominate the spectrum of the 100 keV situation, while the photons that are scattered 3-5 times dominate the spectrum in the 200 keV situation. As seen in Fig.~\ref{fig:polsca_corona}, for the same inclination and the number of scatterings, the polarization degrees from the slab corona are close for two different electron temperature situations. With a fixed photon index, higher temperature requires lower optical depth, leading to less scatterings reducing the X-ray polarization degree.

\subsection{Jet}

\begin{figure}
	\includegraphics[width=\columnwidth]{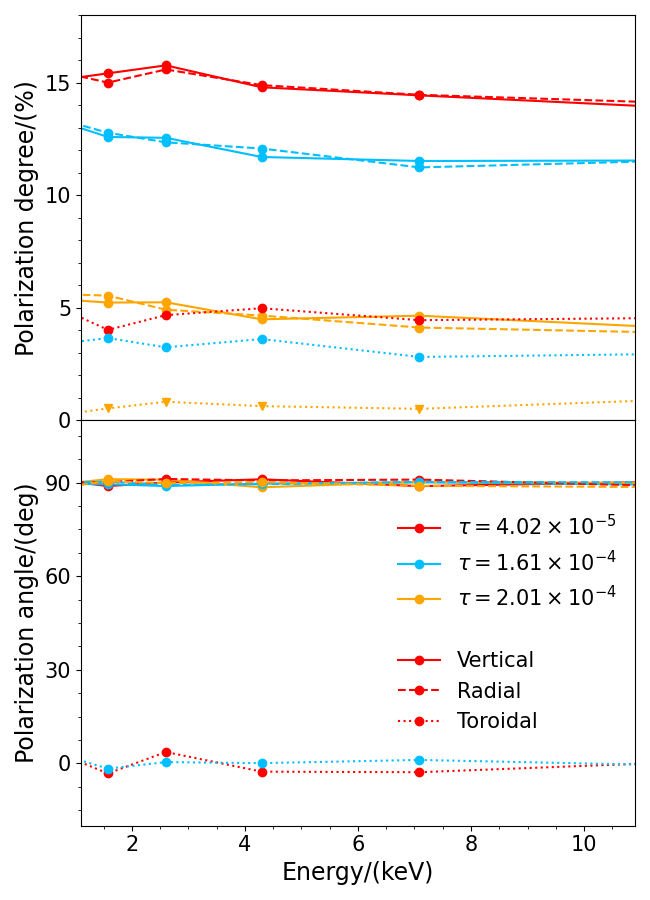}
    \caption{The 2-10 keV polarization degree and polarization angle of the jet. The vertical, radial and toroidal magnetic field situations are shown as solid, dashed, and dotted line. The optical depth of these situations are marked by color: red, blue, and yellow for $\tau=4.02\times10^{-5}$, 1.61$\times10^{-4}$, and $2.01\times10^{-4}$, respectively. The values of polarization degree are marked by solid circle. The upper limits of polarization degree are marked by solid triangle. All standard errors of polarization degree  in this band for jet are below 0.48\%. All standard errors of polarization angle in this band for jet are below 2.6$^{\circ}$.}
    \label{fig:pol_jet}
\end{figure}

The polarization degree and angle of the jet in 2-10 keV are shown in Fig.~\ref{fig:pol_jet}. The magnetic field configuration has a significant impact on the polarization degree of the SSC process. The jet with {\color{red}\textbf{a}} vertical or radial field has higher polarization degrees (4.1\%-15.8\%) than that of the toroidal field ($\leq$5.0\%) because the angle between the line of sight and the magnetic field direction is smaller for the vertical and radial field than the angle for the toroidal field. Since the opening angle of the jet is only 2.8$^{\circ}$, the results of the vertical and radial fields are similar. Different from the corona, the polarization degree of the jet decreases with the optical depth. For the vertical and radial fields, the polarization degrees of the jet are around 14.4\%-15.8\%, 11.2\%-12.8\%, and 4.1\%-5.5\% for $\tau=4.02\times10^{-5}$ ($\emph{N}_{\rm e}=$ 10 cm$^{-3}$), $ 1.61\times10^{-4}$ ($\emph{N}_{\rm e}=$ 100 cm$^{-3}$), and $2.01\times10^{-4}$ ($\emph{N}_{\rm e}=$ 50 cm$^{-3}$), respectively. For the toroidal field, the polarization degrees are around 4.0\%-5.0\%, and 2.8\%-3.7\% for $\tau=4.02\times10^{-5}$ ($\emph{N}_{\rm e}=$ 10 cm$^{-3}$), and $ 1.61\times10^{-4}$ ($\emph{N}_{\rm e}=$ 100 cm$^{-3}$), respectively. These are larger than the polarization degrees of the sphere and slab coronae. For the toroidal field with $\tau=2.01\times10^{-4}$ ($\emph{N}_{\rm e}=$ 50 cm$^{-3}$), the radiation in 2-10 keV is also unpolarized. In addition, the polarization degrees of the jet slightly decrease with the photon energy.

The polarization angle of the jet also depends mainly on the magnetic field direction. For a jet with a vertical or radial field, the polarization is perpendicular to the direction of the jet. For the toroidal field, the polarization is parallel to the direction of the jet for $\tau=4.02\times10^{-5}$ ($\emph{N}_{\rm e}=$ 10 cm$^{-3}$) and $1.61\times10^{-4}$ ($\emph{N}_{\rm e}=$ 100 cm$^{-3}$).

\begin{figure*}
	\includegraphics[width=\textwidth]{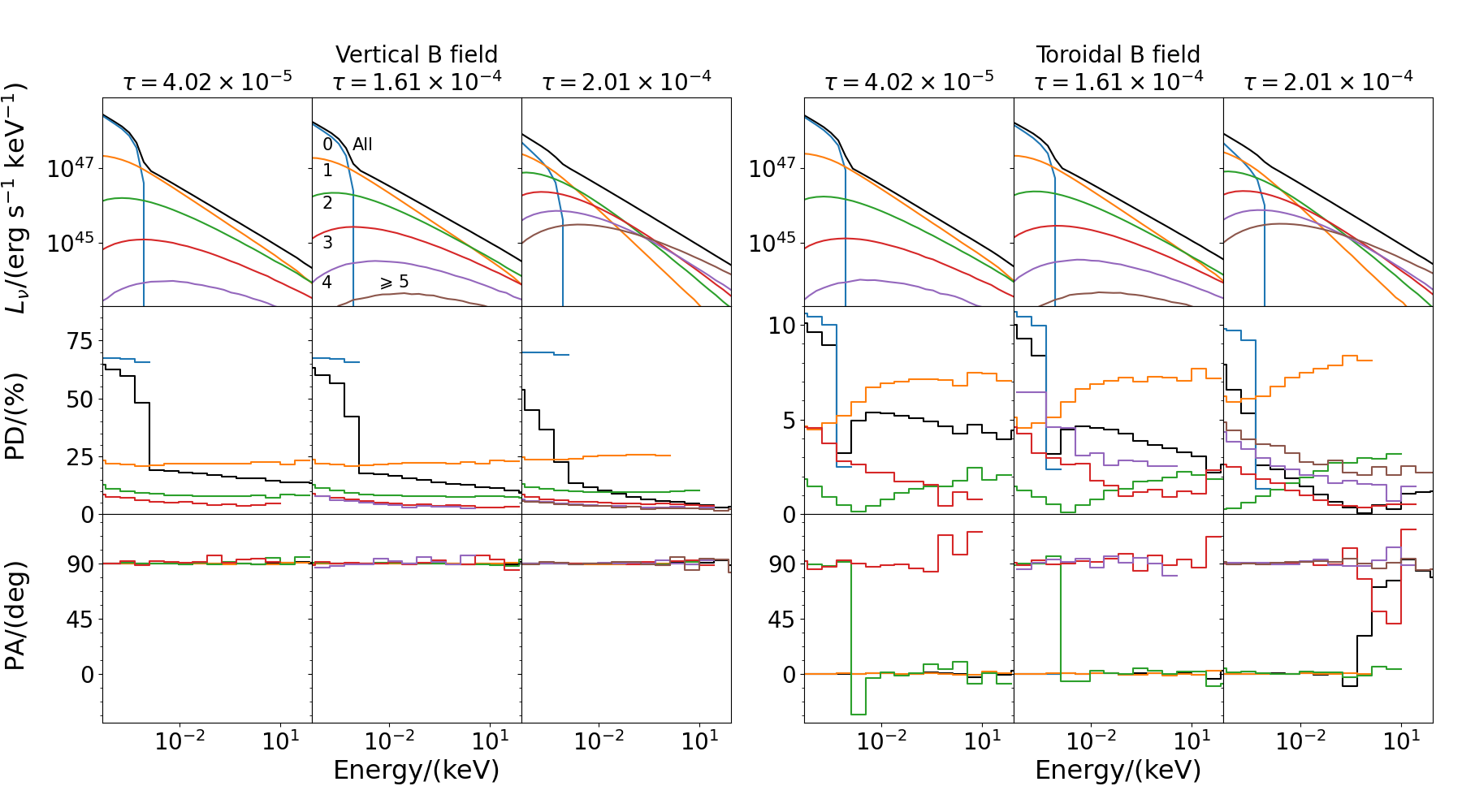}
    \caption{The luminosity ($L_\nu$), polarization degree (PD), and polarization angle (PA) of the jet in vertical (left 3 columns) and toroidal (right 3 columns) magnetic field for different numbers of scatterings. The photons which are scattered 0, 1, 2, 3, 4, and 5 or more times are represented by blue, yellow, green, red, purple, and brown lines, respectively, while the results of total emission are represented by black lines.}
    \label{fig:polsca_bv_bt}
\end{figure*}

To investigate the effects of the optical depth on the X-ray polarization of the jet, we also calculate the polarization degrees and angles of the jet at several different numbers of scatterings. In Fig.~\ref{fig:polsca_bv_bt}, optical depth changes the dominant number of scatterings to 2-10 keV. The photons that are scattered once, 1-2 times, and 3-4 times dominate the energy spectrum in 2-10 keV for $\tau=4.02\times10^{-5}$ ($\emph{N}_{\rm e}=$ 10 cm$^{-3}$), $1.61\times10^{-4}$ ($\emph{N}_{\rm e}=$ 100 cm$^{-3}$), and $2.01\times10^{-4}$ ($\emph{N}_{\rm e}=$ 50 cm$^{-3}$), respectively. Unlike the corona, the seed photons of the jet are highly polarized. For the SSC process, the seed photons from synchrotron radiation have high polarization degrees. In each inverse Compton scattering, the polarization orientations of highly polarized photons are gradually more isotropic in each scattering given the isotropic electron velocity distribution. Therefore, the SSC radiation from the jet with a larger optical depth has a lower polarization degree. Due to the relationship between magnetic field and polarization angle orientation, the synchrotron radiations from vertical and toroidal fields have different polarization angles. The photons that are scattered once tend to keep the same polarization angles as the seed photons from synchrotron radiation, while the multi-scattered photons tend to have polarization perpendicular to the jet. Therefore, the optical depth also influences the polarization angle of the jet with the toroidal magnetic field.

\section{Discussion}
\label{sec:dis}

Our results show that the polarization can effectively distinguish the geometrical and physical parameters of the corona and jet. Here, we summarize the main polarization characteristics of the jet and corona for 3C 273. The polarization degrees in 2-10 keV from both the sphere and slab coronae are unpolarized, while these are 4.1\%-15.8\% from the jet with a vertical or radial magnetic fields, 2.8\%-5.0\% for the jet with toroidal magnetic field and the optical depth below $1.61\times10^{-4}$, and almost unpolarized for the jet with toroidal magnetic field and optical depth around $2.01\times10^{-4}$. Furthermore, other X-ray components, such as reflection, should not qualitatively influence these results due to the low reflection fractions in X-ray ($R=0.02-0.2$) of 3C 273 \citep{2015ApJ...812...14M}.

The new X-ray polarimetric observatory, \emph{IXPE} \citep{2016SPIE.9905E..17W}, provides the opportunity for high-sensitivity soft X-ray polarization observations. The sensitivity of \emph{IXPE} is expressed by the minimum detectable polarization (MDP) at 99\% confidence level. We calculate the MDP$_{99}$ by WebPIMMS\footnote{https://heasarc.gsfc.nasa.gov/cgi-bin/Tools/w3pimms/w3pimms.pl}. For 3C 273, with photon index taken as 1.59, flux in 2.5-10 keV taken as 8.8$\times$10$^{-11}$ erg cm$^{-2}$ s$^{-1}$ and neutral H column density taken as 1.8$\times10^{-20}$ cm$^{-2}$ \citep{1990ARA&A..28..215D}, the exposure time to achieve MDP$_{99}$ of 2\% in 2-8 keV for \emph{IXPE} is around 890 ks. Up to now, the upper limit of 9\% to X-ray polarization degree from the 95.28 ks observation by \emph{IXPE} \citep{2023arXiv231011510M} has eliminated the jet in vertical and radial magnetic field dominating situations. The future polarimetric observation with longer exposure time can better constrain the X-ray radiation origin of 3C 273.

\section{Conclusions}
\label{sec:con}

This work investigates the polarization properties of the sphere corona, slab corona, and jet. The results of our work can be used to constrain the X-ray radiation origin and the geometrical and physical properties of 3C 273 with X-ray polarimetric observations taken by, e.g., \emph{IXPE}. The main spectral and polarization properties of the corona and jet are summarized below.

We set geometrical and physical parameters for the corona and jet in our simulations by fitting the observed flux and spectra shape in 2.5-10 keV of 3C 273 in the low state \citep{2017MNRAS.469.3824K}. We find that sphere coronae with radii of 3.7-12 \emph{R}$_{\rm G}$ and partially covering slab coronae can reproduce the observed energy spectra for any observer’s inclination, while jets can reproduce the observed energy spectra if the observer is located at the inclination of 0$^{\circ}$-12$^{\circ}$.

\begin{figure*}
	\includegraphics[width=\textwidth]{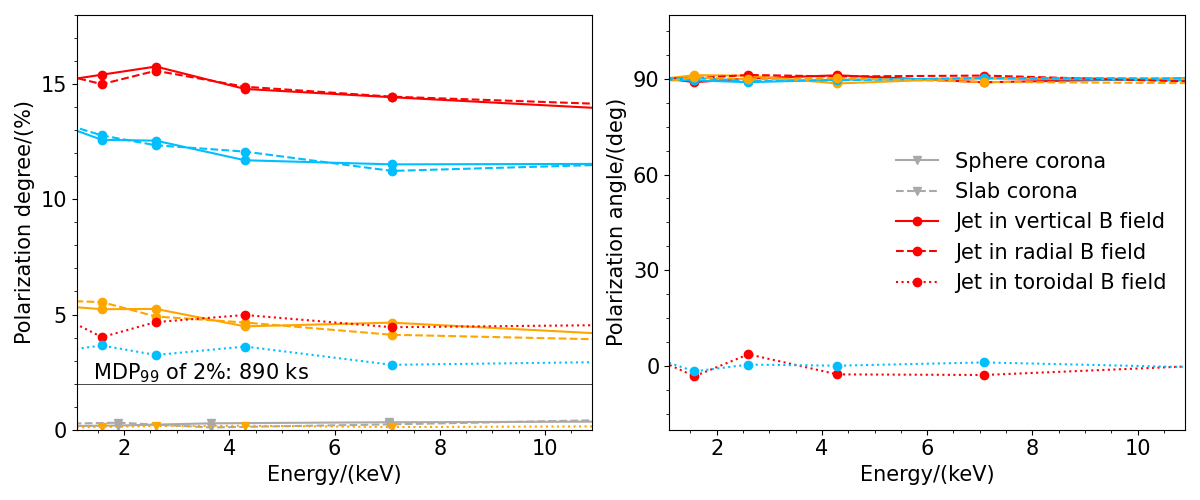}
    \caption{The polarization of 3C 273 for different corona and jet. The results of sphere corona are the ones whose electron temperature is 200 keV, black hole spin is 0.998, and $h=20$ $\emph{R}_{\rm G}$, while the results of slab corona are also the ones with 200 keV electron temperature, and Kerr black hole. The various optical depths of jets are marked by different colors: red, blue, and yellow for $\tau=4.02\times10^{-5}$, 1.61$\times10^{-4}$, and $2.01\times10^{-4}$, respectively. The values of polarization degree are marked by solid circle. The upper limits of polarization degree are marked by solid triangle.}
    \label{fig:pol_summary}
\end{figure*}

Our major finding is that the differences in X-ray polarization between the jet and the corona are significant. This shows that X-ray polarization can be used to constrain the origin of the X-ray radiation of 3C 273 in the low state. We expect that the X-ray radiations in 2-10 keV from both the sphere and slab coronae of 3C 273 are almost unpolarized, while the polarization degrees of the jet are much larger: 4.1\%-15.8\% for the jet with a vertical or radial magnetic field, 2.8\%-5.0\% for the jet with toroidal magnetic field and the optical depth below $1.61\times10^{-4}$, and unpolarized for the jet with toroidal magnetic field and optical depth around $2.01\times10^{-4}$. The polarization angles are always perpendicular to the direction of the black hole rotation axis for the jet in a vertical or radial magnetic field, while they are parallel to the direction of the black hole
rotation axis for the jet with a toroidal magnetic field if the optical depth is lower than $1.61\times10^{-4}$ and perpendicular for larger optical depth. The main results of our simulations are summarized in Fig.~\ref{fig:pol_summary}. The current observation by \emph{IXPE} does not support the jet-dominating model in a vertical or radial magnetic field.

We also find that the X-ray polarization of the corona and jet is sensitive to optical depth and geometry, and the main driver for this dependence is the number of scatterings.

Finally, to better determine the origin of the X-ray radiation of 3C 273 in the low state in the future, MDP$_{99}$ needs to arrive at 2\%. The corresponding exposure time of 3C 273 is around 890 ks for \emph{IXPE}.

\section*{Acknowledgements}

We thank the anonymous referee for his/her useful comments that greatly improved the manuscript. We acknowledge the support by the Strategic Pioneer Program on Space Science, Chinese Academy of Sciences through grants XDA15310000, XDA15052100, and the support by the Strategic Priority Research Program of the Chinese Academy of Sciences, Grant No. XDB0550200. This work is also supported by the National Natural Science Foundation of China (grant 12073037).

\section*{Data Availability}

The data underlying this article will be shared on reasonable request to the corresponding author.




\begin{thebibliography}{}
\makeatletter
\relax
\def\mn@urlcharsother{\let\do\@makeother \do\$\do\&\do\#\do\^\do\_\do\%\do\~}
\def\mn@doi{\begingroup\mn@urlcharsother \@ifnextchar [ {\mn@doi@}
  {\mn@doi@[]}}
\def\mn@doi@[#1]#2{\def\@tempa{#1}\ifx\@tempa\@empty \href
  {http://dx.doi.org/#2} {doi:#2}\else \href {http://dx.doi.org/#2} {#1}\fi
  \endgroup}
\def\mn@eprint#1#2{\mn@eprint@#1:#2::\@nil}
\def\mn@eprint@arXiv#1{\href {http://arxiv.org/abs/#1} {{\tt arXiv:#1}}}
\def\mn@eprint@dblp#1{\href {http://dblp.uni-trier.de/rec/bibtex/#1.xml}
  {dblp:#1}}
\def\mn@eprint@#1:#2:#3:#4\@nil{\def\@tempa {#1}\def\@tempb {#2}\def\@tempc
  {#3}\ifx \@tempc \@empty \let \@tempc \@tempb \let \@tempb \@tempa \fi \ifx
  \@tempb \@empty \def\@tempb {arXiv}\fi \@ifundefined
  {mn@eprint@\@tempb}{\@tempb:\@tempc}{\expandafter \expandafter \csname
  mn@eprint@\@tempb\endcsname \expandafter{\@tempc}}}

\bibitem[\protect\citeauthoryear{{Appenzeller}}{{Appenzeller}}{1968}]{1968ApJ...151..769A}
{Appenzeller} I.,  1968, \mn@doi [\apj] {10.1086/149476}, \href
  {https://ui.adsabs.harvard.edu/abs/1968ApJ...151..769A} {151, 769}

\bibitem[\protect\citeauthoryear{{Attridge}, {Wardle}  \& {Homan}}{{Attridge}
  et~al.}{2005}]{2005ApJ...633L..85A}
{Attridge} J.~M.,  {Wardle} J. F.~C.,   {Homan} D.~C.,  2005, \mn@doi [\apjl]
  {10.1086/498392}, \href
  {https://ui.adsabs.harvard.edu/abs/2005ApJ...633L..85A} {633, L85}

\bibitem[\protect\citeauthoryear{{Cappi}, {Matsuoka}, {Otani}  \&
  {Leighly}}{{Cappi} et~al.}{1998}]{1998PASJ...50..213C}
{Cappi} M.,  {Matsuoka} M.,  {Otani} C.,   {Leighly} K.~M.,  1998, \mn@doi
  [\pasj] {10.1093/pasj/50.2.213}, \href
  {https://ui.adsabs.harvard.edu/abs/1998PASJ...50..213C} {50, 213}

\bibitem[\protect\citeauthoryear{{Chidiac} et~al.,}{{Chidiac}
  et~al.}{2016}]{2016A&A...590A..61C}
{Chidiac} C.,  et~al., 2016, \mn@doi [\aap] {10.1051/0004-6361/201628347},
  \href {https://ui.adsabs.harvard.edu/abs/2016A&A...590A..61C} {590, A61}

\bibitem[\protect\citeauthoryear{{Conway} \& {Davis}}{{Conway} \&
  {Davis}}{1994}]{1994A&A...284..724C}
{Conway} R.~G.,  {Davis} R.~J.,  1994, \aap, \href
  {https://ui.adsabs.harvard.edu/abs/1994A&A...284..724C} {284, 724}

\bibitem[\protect\citeauthoryear{{Conway}, {Garrington}, {Perley}  \&
  {Biretta}}{{Conway} et~al.}{1993}]{1993A&A...267..347C}
{Conway} R.~G.,  {Garrington} S.~T.,  {Perley} R.~A.,   {Biretta} J.~A.,  1993,
  \aap, \href {https://ui.adsabs.harvard.edu/abs/1993A&A...267..347C} {267,
  347}

\bibitem[\protect\citeauthoryear{{Courvoisier}}{{Courvoisier}}{1998}]{1998A&ARv...9....1C}
{Courvoisier} T. J.~L.,  1998, \mn@doi [\aapr] {10.1007/s001590050013}, \href
  {https://ui.adsabs.harvard.edu/abs/1998A&ARv...9....1C} {9, 1}

\bibitem[\protect\citeauthoryear{{Deluit}, {Stuhlinger}  \&
  {Staubert}}{{Deluit} et~al.}{2006}]{2006AdSpR..38.1393D}
{Deluit} S.~J.,  {Stuhlinger} M.,   {Staubert} R.,  2006, \mn@doi [Advances in
  Space Research] {10.1016/j.asr.2005.05.090}, \href
  {https://ui.adsabs.harvard.edu/abs/2006AdSpR..38.1393D} {38, 1393}

\bibitem[\protect\citeauthoryear{{Di Gesu} et~al.,}{{Di Gesu}
  et~al.}{2022}]{2022ApJ...938L...7D}
{Di Gesu} L.,  et~al., 2022, \mn@doi [\apjl] {10.3847/2041-8213/ac913a}, \href
  {https://ui.adsabs.harvard.edu/abs/2022ApJ...938L...7D} {938, L7}

\bibitem[\protect\citeauthoryear{{Dickey} \& {Lockman}}{{Dickey} \&
  {Lockman}}{1990}]{1990ARA&A..28..215D}
{Dickey} J.~M.,  {Lockman} F.~J.,  1990, \mn@doi [\araa]
  {10.1146/annurev.aa.28.090190.001243}, \href
  {https://ui.adsabs.harvard.edu/abs/1990ARA&A..28..215D} {28, 215}

\bibitem[\protect\citeauthoryear{{Done}}{{Done}}{2014}]{2014xru..confE..64D}
{Done} C.,  2014, in {Ness} J.-U.,  ed., The X-ray Universe 2014. p.~64

\bibitem[\protect\citeauthoryear{{Esposito}, {Walter}, {Jean}, {Tramacere},
  {T{\"u}rler}, {L{\"a}hteenm{\"a}ki}  \& {Tornikoski}}{{Esposito}
  et~al.}{2015}]{2015A&A...576A.122E}
{Esposito} V.,  {Walter} R.,  {Jean} P.,  {Tramacere} A.,  {T{\"u}rler} M.,
  {L{\"a}hteenm{\"a}ki} A.,   {Tornikoski} M.,  2015, \mn@doi [\aap]
  {10.1051/0004-6361/201424644}, \href
  {https://ui.adsabs.harvard.edu/abs/2015A&A...576A.122E} {576, A122}

\bibitem[\protect\citeauthoryear{{Gianolli} et~al.,}{{Gianolli}
  et~al.}{2023}]{2023MNRAS.523.4468G}
{Gianolli} V.~E.,  et~al., 2023, \mn@doi [\mnras] {10.1093/mnras/stad1697},
  \href {https://ui.adsabs.harvard.edu/abs/2023MNRAS.523.4468G} {523, 4468}

\bibitem[\protect\citeauthoryear{{Gnarini}, {Ursini}, {Matt}, {Bianchi},
  {Capitanio}, {Cocchi}, {Farinelli}  \& {Zhang}}{{Gnarini}
  et~al.}{2022}]{2022MNRAS.514.2561G}
{Gnarini} A.,  {Ursini} F.,  {Matt} G.,  {Bianchi} S.,  {Capitanio} F.,
  {Cocchi} M.,  {Farinelli} R.,   {Zhang} W.,  2022, \mn@doi [\mnras]
  {10.1093/mnras/stac1523}, \href
  {https://ui.adsabs.harvard.edu/abs/2022MNRAS.514.2561G} {514, 2561}

\bibitem[\protect\citeauthoryear{{Goosmann} \& {Gaskell}}{{Goosmann} \&
  {Gaskell}}{2007}]{2007A&A...465..129G}
{Goosmann} R.~W.,  {Gaskell} C.~M.,  2007, \mn@doi [\aap]
  {10.1051/0004-6361:20053555}, \href
  {https://ui.adsabs.harvard.edu/abs/2007A&A...465..129G} {465, 129}

\bibitem[\protect\citeauthoryear{{Grandi} \& {Palumbo}}{{Grandi} \&
  {Palumbo}}{2004}]{2004Sci...306..998G}
{Grandi} P.,  {Palumbo} G. G.~C.,  2004, \mn@doi [Science]
  {10.1126/science.1101787}, \href
  {https://ui.adsabs.harvard.edu/abs/2004Sci...306..998G} {306, 998}

\bibitem[\protect\citeauthoryear{{Haardt} \& {Maraschi}}{{Haardt} \&
  {Maraschi}}{1991}]{1991ApJ...380L..51H}
{Haardt} F.,  {Maraschi} L.,  1991, \mn@doi [\apjl] {10.1086/186171}, \href
  {https://ui.adsabs.harvard.edu/abs/1991ApJ...380L..51H} {380, L51}

\bibitem[\protect\citeauthoryear{{Haardt} et~al.,}{{Haardt}
  et~al.}{1998}]{1998A&A...340...35H}
{Haardt} F.,  et~al., 1998, \aap, \href
  {https://ui.adsabs.harvard.edu/abs/1998A&A...340...35H} {340, 35}

\bibitem[\protect\citeauthoryear{{Hada} et~al.,}{{Hada}
  et~al.}{2016}]{2016ApJ...817..131H}
{Hada} K.,  et~al., 2016, \mn@doi [\apj] {10.3847/0004-637X/817/2/131}, \href
  {https://ui.adsabs.harvard.edu/abs/2016ApJ...817..131H} {817, 131}

\bibitem[\protect\citeauthoryear{{Haddock} \& {Hobbs}}{{Haddock} \&
  {Hobbs}}{1963}]{1963AJ.....68Q..75H}
{Haddock} F.~T.,  {Hobbs} R.~W.,  1963, \mn@doi [\aj] {10.1086/109021}, \href
  {https://ui.adsabs.harvard.edu/abs/1963AJ.....68Q..75H} {68, 75}

\bibitem[\protect\citeauthoryear{{Hinshaw} et~al.,}{{Hinshaw}
  et~al.}{2013}]{2013ApJS..208...19H}
{Hinshaw} G.,  et~al., 2013, \mn@doi [\apjs] {10.1088/0067-0049/208/2/19},
  \href {https://ui.adsabs.harvard.edu/abs/2013ApJS..208...19H} {208, 19}

\bibitem[\protect\citeauthoryear{{Homan}, {Ojha}, {Wardle}, {Roberts}, {Aller},
  {Aller}  \& {Hughes}}{{Homan} et~al.}{2002}]{2002ApJ...568...99H}
{Homan} D.~C.,  {Ojha} R.,  {Wardle} J. F.~C.,  {Roberts} D.~H.,  {Aller}
  M.~F.,  {Aller} H.~D.,   {Hughes} P.~A.,  2002, \mn@doi [\apj]
  {10.1086/338701}, \href
  {https://ui.adsabs.harvard.edu/abs/2002ApJ...568...99H} {568, 99}

\bibitem[\protect\citeauthoryear{{Hovatta}, {O'Sullivan}, {Mart{\'\i}-Vidal},
  {Savolainen}  \& {Tchekhovskoy}}{{Hovatta}
  et~al.}{2019}]{2019A&A...623A.111H}
{Hovatta} T.,  {O'Sullivan} S.,  {Mart{\'\i}-Vidal} I.,  {Savolainen} T.,
  {Tchekhovskoy} A.,  2019, \mn@doi [\aap] {10.1051/0004-6361/201832587}, \href
  {https://ui.adsabs.harvard.edu/abs/2019A&A...623A.111H} {623, A111}

\bibitem[\protect\citeauthoryear{{Impey}, {Malkan}  \& {Tapia}}{{Impey}
  et~al.}{1989}]{1989ApJ...347...96I}
{Impey} C.~D.,  {Malkan} M.~A.,   {Tapia} S.,  1989, \mn@doi [\apj]
  {10.1086/168100}, \href
  {https://ui.adsabs.harvard.edu/abs/1989ApJ...347...96I} {347, 96}

\bibitem[\protect\citeauthoryear{{Ingram} et~al.,}{{Ingram}
  et~al.}{2023a}]{2023arXiv230513028I}
{Ingram} A.,  et~al., 2023a, \mn@doi [arXiv e-prints]
  {10.48550/arXiv.2305.13028}, \href
  {https://ui.adsabs.harvard.edu/abs/2023arXiv230513028I} {p. arXiv:2305.13028}

\bibitem[\protect\citeauthoryear{{Ingram} et~al.,}{{Ingram}
  et~al.}{2023b}]{2023MNRAS.525.5437I}
{Ingram} A.,  et~al., 2023b, \mn@doi [\mnras] {10.1093/mnras/stad2625}, \href
  {https://ui.adsabs.harvard.edu/abs/2023MNRAS.525.5437I} {525, 5437}

\bibitem[\protect\citeauthoryear{{Jolley}, {Kuncic}, {Bicknell}  \&
  {Wagner}}{{Jolley} et~al.}{2009}]{2009MNRAS.400.1521J}
{Jolley} E.~J.~D.,  {Kuncic} Z.,  {Bicknell} G.~V.,   {Wagner} S.,  2009,
  \mn@doi [\mnras] {10.1111/j.1365-2966.2009.15554.x}, \href
  {https://ui.adsabs.harvard.edu/abs/2009MNRAS.400.1521J} {400, 1521}

\bibitem[\protect\citeauthoryear{{Jorstad} et~al.,}{{Jorstad}
  et~al.}{2005a}]{2005AJ....130.1418J}
{Jorstad} S.~G.,  et~al., 2005a, \mn@doi [\aj] {10.1086/444593}, \href
  {https://ui.adsabs.harvard.edu/abs/2005AJ....130.1418J} {130, 1418}

\bibitem[\protect\citeauthoryear{{Jorstad} et~al.,}{{Jorstad}
  et~al.}{2005b}]{2005ASPC..340..183J}
{Jorstad} S.,  et~al., 2005b, in {Romney} J.,  {Reid} M.,  eds,  Astronomical
  Society of the Pacific Conference Series Vol. 340, Future Directions in High
  Resolution Astronomy. p.~183

\bibitem[\protect\citeauthoryear{{Jorstad} et~al.,}{{Jorstad}
  et~al.}{2007}]{2007AJ....134..799J}
{Jorstad} S.~G.,  et~al., 2007, \mn@doi [\aj] {10.1086/519996}, \href
  {https://ui.adsabs.harvard.edu/abs/2007AJ....134..799J} {134, 799}

\bibitem[\protect\citeauthoryear{{Jorstad} et~al.,}{{Jorstad}
  et~al.}{2017}]{2017ApJ...846...98J}
{Jorstad} S.~G.,  et~al., 2017, \mn@doi [\apj] {10.3847/1538-4357/aa8407},
  \href {https://ui.adsabs.harvard.edu/abs/2017ApJ...846...98J} {846, 98}

\bibitem[\protect\citeauthoryear{{Kalita}, {Gupta}, {Wiita}, {Dewangan}  \&
  {Duorah}}{{Kalita} et~al.}{2017}]{2017MNRAS.469.3824K}
{Kalita} N.,  {Gupta} A.~C.,  {Wiita} P.~J.,  {Dewangan} G.~C.,   {Duorah} K.,
  2017, \mn@doi [\mnras] {10.1093/mnras/stx1108}, \href
  {https://ui.adsabs.harvard.edu/abs/2017MNRAS.469.3824K} {469, 3824}

\bibitem[\protect\citeauthoryear{{Kemp}, {Rieke}, {Lebofsky}  \&
  {Coyne}}{{Kemp} et~al.}{1977}]{1977ApJ...215L.107K}
{Kemp} J.~C.,  {Rieke} G.~H.,  {Lebofsky} M.~J.,   {Coyne} G.~V.,  1977,
  \mn@doi [\apjl] {10.1086/182490}, \href
  {https://ui.adsabs.harvard.edu/abs/1977ApJ...215L.107K} {215, L107}

\bibitem[\protect\citeauthoryear{{Kim}, {Trippe}  \& {Kravchenko}}{{Kim}
  et~al.}{2020}]{2020A&A...636A..62K}
{Kim} D.-W.,  {Trippe} S.,   {Kravchenko} E.~V.,  2020, \mn@doi [\aap]
  {10.1051/0004-6361/202037474}, \href
  {https://ui.adsabs.harvard.edu/abs/2020A&A...636A..62K} {636, A62}

\bibitem[\protect\citeauthoryear{{Knacke}, {Johns}  \& {Capps}}{{Knacke}
  et~al.}{1979}]{1979Natur.280..215K}
{Knacke} R.~F.,  {Johns} M.,   {Capps} R.~W.,  1979, \mn@doi [\nat]
  {10.1038/280215a0}, \href
  {https://ui.adsabs.harvard.edu/abs/1979Natur.280..215K} {280, 215}

\bibitem[\protect\citeauthoryear{{Leung}, {Gammie}  \& {Noble}}{{Leung}
  et~al.}{2011}]{2011ApJ...737...21L}
{Leung} P.~K.,  {Gammie} C.~F.,   {Noble} S.~C.,  2011, \mn@doi [\apj]
  {10.1088/0004-637X/737/1/21}, \href
  {https://ui.adsabs.harvard.edu/abs/2011ApJ...737...21L} {737, 21}

\bibitem[\protect\citeauthoryear{{Li}, {Zhang}, {Jin}, {Du}, {Cui}, {Liu}  \&
  {Wang}}{{Li} et~al.}{2020}]{2020ApJ...897...18L}
{Li} Y.-R.,  {Zhang} Z.-X.,  {Jin} C.,  {Du} P.,  {Cui} L.,  {Liu} X.,   {Wang}
  J.-M.,  2020, \mn@doi [\apj] {10.3847/1538-4357/ab95a3}, \href
  {https://ui.adsabs.harvard.edu/abs/2020ApJ...897...18L} {897, 18}

\bibitem[\protect\citeauthoryear{{Li}, {Wang}, {Songsheng}, {Zhang}, {Du}, {Hu}
   \& {Xiao}}{{Li} et~al.}{2022}]{2022ApJ...927...58L}
{Li} Y.-R.,  {Wang} J.-M.,  {Songsheng} Y.-Y.,  {Zhang} Z.-X.,  {Du} P.,  {Hu}
  C.,   {Xiao} M.,  2022, \mn@doi [\apj] {10.3847/1538-4357/ac4bcb}, \href
  {https://ui.adsabs.harvard.edu/abs/2022ApJ...927...58L} {927, 58}

\bibitem[\protect\citeauthoryear{{Liller}}{{Liller}}{1969}]{1969ApJ...155.1113L}
{Liller} W.,  1969, \mn@doi [\apj] {10.1086/149937}, \href
  {https://ui.adsabs.harvard.edu/abs/1969ApJ...155.1113L} {155, 1113}

\bibitem[\protect\citeauthoryear{{Lisakov}, {Kovalev}, {Savolainen}, {Hovatta}
  \& {Kutkin}}{{Lisakov} et~al.}{2017}]{2017MNRAS.468.4478L}
{Lisakov} M.~M.,  {Kovalev} Y.~Y.,  {Savolainen} T.,  {Hovatta} T.,   {Kutkin}
  A.~M.,  2017, \mn@doi [\mnras] {10.1093/mnras/stx710}, \href
  {https://ui.adsabs.harvard.edu/abs/2017MNRAS.468.4478L} {468, 4478}

\bibitem[\protect\citeauthoryear{{Madsen} et~al.,}{{Madsen}
  et~al.}{2015}]{2015ApJ...812...14M}
{Madsen} K.~K.,  et~al., 2015, \mn@doi [\apj] {10.1088/0004-637X/812/1/14},
  \href {https://ui.adsabs.harvard.edu/abs/2015ApJ...812...14M} {812, 14}

\bibitem[\protect\citeauthoryear{{Marin} \& {Weisskopf}}{{Marin} \&
  {Weisskopf}}{2017}]{2017sf2a.conf..173M}
{Marin} F.,  {Weisskopf} M.~C.,  2017, in {Reyl{\'e}} C.,  {Di Matteo} P.,
  {Herpin} F.,  {Lagadec} E.,  {Lan{\c{c}}on} A.,  {Meliani} Z.,   {Royer} F.,
  eds, SF2A-2017: Proceedings of the Annual meeting of the French Society of
  Astronomy and Astrophysics. p.~Di (\mn@eprint {arXiv} {1708.02022})

\bibitem[\protect\citeauthoryear{{Marinucci} et~al.,}{{Marinucci}
  et~al.}{2022}]{2022arXiv220709338M}
{Marinucci} A.,  et~al., 2022, arXiv e-prints, \href
  {https://ui.adsabs.harvard.edu/abs/2022arXiv220709338M} {p. arXiv:2207.09338}

\bibitem[\protect\citeauthoryear{{Marshall} et~al.,}{{Marshall}
  et~al.}{2023}]{2023arXiv231011510M}
{Marshall} H.~L.,  et~al., 2023, \mn@doi [arXiv e-prints]
  {10.48550/arXiv.2310.11510}, \href
  {https://ui.adsabs.harvard.edu/abs/2023arXiv231011510M} {p. arXiv:2310.11510}

\bibitem[\protect\citeauthoryear{{Meyer} \& {Georganopoulos}}{{Meyer} \&
  {Georganopoulos}}{2014}]{2014ApJ...780L..27M}
{Meyer} E.~T.,  {Georganopoulos} M.,  2014, \mn@doi [\apjl]
  {10.1088/2041-8205/780/2/L27}, \href
  {https://ui.adsabs.harvard.edu/abs/2014ApJ...780L..27M} {780, L27}

\bibitem[\protect\citeauthoryear{{Meyer} et~al.,}{{Meyer}
  et~al.}{2016}]{2016ApJ...818..195M}
{Meyer} E.~T.,  et~al., 2016, \mn@doi [\apj] {10.3847/0004-637X/818/2/195},
  \href {https://ui.adsabs.harvard.edu/abs/2016ApJ...818..195M} {818, 195}

\bibitem[\protect\citeauthoryear{{Moroz} \& {Yessipov}}{{Moroz} \&
  {Yessipov}}{1963}]{1963IBVS...31....1M}
{Moroz} V.~I.,  {Yessipov} V.~F.,  1963, Information Bulletin on Variable
  Stars, \href {https://ui.adsabs.harvard.edu/abs/1963IBVS...31....1M} {31, 1}

\bibitem[\protect\citeauthoryear{{Morrison} \& {Sadun}}{{Morrison} \&
  {Sadun}}{1992}]{1992MNRAS.254..488M}
{Morrison} P.,  {Sadun} A.,  1992, \mn@doi [\mnras] {10.1093/mnras/254.3.488},
  \href {https://ui.adsabs.harvard.edu/abs/1992MNRAS.254..488M} {254, 488}

\bibitem[\protect\citeauthoryear{{Narayan} et~al.,}{{Narayan}
  et~al.}{2021}]{2021ApJ...912...35N}
{Narayan} R.,  et~al., 2021, \mn@doi [\apj] {10.3847/1538-4357/abf117}, \href
  {https://ui.adsabs.harvard.edu/abs/2021ApJ...912...35N} {912, 35}

\bibitem[\protect\citeauthoryear{{Novikov} \& {Thorne}}{{Novikov} \&
  {Thorne}}{1973}]{1973blho.conf..343N}
{Novikov} I.~D.,  {Thorne} K.~S.,  1973, in Black Holes (Les Astres Occlus). pp
  343--450

\bibitem[\protect\citeauthoryear{{Park} \& {Algaba}}{{Park} \&
  {Algaba}}{2022}]{2022arXiv221013819P}
{Park} J.,  {Algaba} J.~C.,  2022, arXiv e-prints, \href
  {https://ui.adsabs.harvard.edu/abs/2022arXiv221013819P} {p. arXiv:2210.13819}

\bibitem[\protect\citeauthoryear{{Pearson}, {Unwin}, {Cohen}, {Linfield},
  {Readhead}, {Seielstad}, {Simon}  \& {Walker}}{{Pearson}
  et~al.}{1981}]{1981Natur.290..365P}
{Pearson} T.~J.,  {Unwin} S.~C.,  {Cohen} M.~H.,  {Linfield} R.~P.,  {Readhead}
  A.~C.~S.,  {Seielstad} G.~A.,  {Simon} R.~S.,   {Walker} R.~C.,  1981,
  \mn@doi [\nat] {10.1038/290365a0}, \href
  {https://ui.adsabs.harvard.edu/abs/1981Natur.290..365P} {290, 365}

\bibitem[\protect\citeauthoryear{{Perley} \& {Meisenheimer}}{{Perley} \&
  {Meisenheimer}}{2017}]{2017A&A...601A..35P}
{Perley} R.~A.,  {Meisenheimer} K.,  2017, \mn@doi [\aap]
  {10.1051/0004-6361/201629704}, \href
  {https://ui.adsabs.harvard.edu/abs/2017A&A...601A..35P} {601, A35}

\bibitem[\protect\citeauthoryear{{Pietrini} \&
  {Torricelli-Ciamponi}}{{Pietrini} \&
  {Torricelli-Ciamponi}}{2008}]{2008A&A...479..365P}
{Pietrini} P.,  {Torricelli-Ciamponi} G.,  2008, \mn@doi [\aap]
  {10.1051/0004-6361:20077597}, \href
  {https://ui.adsabs.harvard.edu/abs/2008A&A...479..365P} {479, 365}

\bibitem[\protect\citeauthoryear{{Podgorn{\'y}}, {Dov{\v{c}}iak}, {Marin},
  {Goosmann}  \& {R{\'o}{\.z}a{\'n}ska}}{{Podgorn{\'y}}
  et~al.}{2022}]{2022MNRAS.510.4723P}
{Podgorn{\'y}} J.,  {Dov{\v{c}}iak} M.,  {Marin} F.,  {Goosmann} R.,
  {R{\'o}{\.z}a{\'n}ska} A.,  2022, \mn@doi [\mnras] {10.1093/mnras/stab3714},
  \href {https://ui.adsabs.harvard.edu/abs/2022MNRAS.510.4723P} {510, 4723}

\bibitem[\protect\citeauthoryear{{Radhakrishnan}, {Seielstad}  \&
  {Berge}}{{Radhakrishnan} et~al.}{1962}]{1962AJ.....67Q.585R}
{Radhakrishnan} V.,  {Seielstad} G.~A.,   {Berge} G.~L.,  1962, \mn@doi [\aj]
  {10.1086/108848}, \href
  {https://ui.adsabs.harvard.edu/abs/1962AJ.....67Q.585R} {67, 585}

\bibitem[\protect\citeauthoryear{{Reis} \& {Miller}}{{Reis} \&
  {Miller}}{2013}]{2013ApJ...769L...7R}
{Reis} R.~C.,  {Miller} J.~M.,  2013, \mn@doi [\apjl]
  {10.1088/2041-8205/769/1/L7}, \href
  {https://ui.adsabs.harvard.edu/abs/2013ApJ...769L...7R} {769, L7}

\bibitem[\protect\citeauthoryear{{Ritacco} et~al.,}{{Ritacco}
  et~al.}{2017}]{2017A&A...599A..34R}
{Ritacco} A.,  et~al., 2017, \mn@doi [\aap] {10.1051/0004-6361/201629666},
  \href {https://ui.adsabs.harvard.edu/abs/2017A&A...599A..34R} {599, A34}

\bibitem[\protect\citeauthoryear{{Roeser} \& {Meisenheimer}}{{Roeser} \&
  {Meisenheimer}}{1986}]{1986A&A...154...15R}
{Roeser} H.~J.,  {Meisenheimer} K.,  1986, \aap, \href
  {https://ui.adsabs.harvard.edu/abs/1986A&A...154...15R} {154, 15}

\bibitem[\protect\citeauthoryear{{Roeser} \& {Meisenheimer}}{{Roeser} \&
  {Meisenheimer}}{1991}]{1991A&A...252..458R}
{Roeser} H.~J.,  {Meisenheimer} K.,  1991, \aap, \href
  {https://ui.adsabs.harvard.edu/abs/1991A&A...252..458R} {252, 458}

\bibitem[\protect\citeauthoryear{{Roeser}, {Conway}  \&
  {Meisenheimer}}{{Roeser} et~al.}{1996}]{1996A&A...314..414R}
{Roeser} H.~J.,  {Conway} R.~G.,   {Meisenheimer} K.,  1996, \aap, \href
  {https://ui.adsabs.harvard.edu/abs/1996A&A...314..414R} {314, 414}

\bibitem[\protect\citeauthoryear{{Rybicki} \& {Lightman}}{{Rybicki} \&
  {Lightman}}{1979}]{1979rpa..book.....R}
{Rybicki} G.~B.,  {Lightman} A.~P.,  1979, {Radiative processes in
  astrophysics}.
A Wiley-Interscience Publication

\bibitem[\protect\citeauthoryear{{Savi{\'c}}, {Goosmann}, {Popovi{\'c}},
  {Marin}  \& {Afanasiev}}{{Savi{\'c}} et~al.}{2018}]{2018A&A...614A.120S}
{Savi{\'c}} D.,  {Goosmann} R.,  {Popovi{\'c}} L.~{\v{C}}.,  {Marin} F.,
  {Afanasiev} V.~L.,  2018, \mn@doi [\aap] {10.1051/0004-6361/201732220}, \href
  {https://ui.adsabs.harvard.edu/abs/2018A&A...614A.120S} {614, A120}

\bibitem[\protect\citeauthoryear{{Savolainen}, {Wiik}, {Valtaoja}  \&
  {Tornikoski}}{{Savolainen} et~al.}{2008}]{2008ASPC..386..451S}
{Savolainen} T.,  {Wiik} K.,  {Valtaoja} E.,   {Tornikoski} M.,  2008, in
  {Rector} T.~A.,  {De Young} D.~S.,  eds,  Astronomical Society of the Pacific
  Conference Series Vol. 386, Extragalactic Jets: Theory and Observation from
  Radio to Gamma Ray. p.~451 (\mn@eprint {arXiv} {0708.0144})

\bibitem[\protect\citeauthoryear{{Scarrott} \& {Warren-Smith}}{{Scarrott} \&
  {Warren-Smith}}{1987}]{1987MNRAS.228P..35S}
{Scarrott} S.~M.,  {Warren-Smith} R.~F.,  1987, \mn@doi [\mnras]
  {10.1093/mnras/228.1.35P}, \href
  {https://ui.adsabs.harvard.edu/abs/1987MNRAS.228P..35S} {228, 35P}

\bibitem[\protect\citeauthoryear{{Schmidt}}{{Schmidt}}{1963}]{1963Natur.197.1040S}
{Schmidt} M.,  1963, \mn@doi [\nat] {10.1038/1971040a0}, \href
  {https://ui.adsabs.harvard.edu/abs/1963Natur.197.1040S} {197, 1040}

\bibitem[\protect\citeauthoryear{{Schmidt}, {Peterson}  \& {Beaver}}{{Schmidt}
  et~al.}{1978}]{1978ApJ...220L..31S}
{Schmidt} G.~D.,  {Peterson} B.~M.,   {Beaver} E.~A.,  1978, \mn@doi [\apjl]
  {10.1086/182631}, \href
  {https://ui.adsabs.harvard.edu/abs/1978ApJ...220L..31S} {220, L31}

\bibitem[\protect\citeauthoryear{{Smith}, {Balonek}, {Elston}  \&
  {Heckert}}{{Smith} et~al.}{1987}]{1987ApJS...64..459S}
{Smith} P.~S.,  {Balonek} T.~J.,  {Elston} R.,   {Heckert} P.~A.,  1987,
  \mn@doi [\apjs] {10.1086/191203}, \href
  {https://ui.adsabs.harvard.edu/abs/1987ApJS...64..459S} {64, 459}

\bibitem[\protect\citeauthoryear{{Soffitta} et~al.,}{{Soffitta}
  et~al.}{2021}]{2021AJ....162..208S}
{Soffitta} P.,  et~al., 2021, \mn@doi [\aj] {10.3847/1538-3881/ac19b0}, \href
  {https://ui.adsabs.harvard.edu/abs/2021AJ....162..208S} {162, 208}

\bibitem[\protect\citeauthoryear{{Soldi} et~al.,}{{Soldi}
  et~al.}{2008}]{2008A&A...486..411S}
{Soldi} S.,  et~al., 2008, \mn@doi [\aap] {10.1051/0004-6361:200809947}, \href
  {https://ui.adsabs.harvard.edu/abs/2008A&A...486..411S} {486, 411}

\bibitem[\protect\citeauthoryear{{Stockman}, {Moore}  \& {Angel}}{{Stockman}
  et~al.}{1984}]{1984ApJ...279..485S}
{Stockman} H.~S.,  {Moore} R.~L.,   {Angel} J.~R.~P.,  1984, \mn@doi [\apj]
  {10.1086/161912}, \href
  {https://ui.adsabs.harvard.edu/abs/1984ApJ...279..485S} {279, 485}

\bibitem[\protect\citeauthoryear{{Sunyaev} \& {Titarchuk}}{{Sunyaev} \&
  {Titarchuk}}{1985}]{1985A&A...143..374S}
{Sunyaev} R.~A.,  {Titarchuk} L.~G.,  1985, \aap, \href
  {https://ui.adsabs.harvard.edu/abs/1985A&A...143..374S} {143, 374}

\bibitem[\protect\citeauthoryear{{Uchiyama} et~al.,}{{Uchiyama}
  et~al.}{2006}]{2006ApJ...648..910U}
{Uchiyama} Y.,  et~al., 2006, \mn@doi [\apj] {10.1086/505964}, \href
  {https://ui.adsabs.harvard.edu/abs/2006ApJ...648..910U} {648, 910}

\bibitem[\protect\citeauthoryear{{Ursini}, {Matt}, {Bianchi}, {Marinucci},
  {Dov{\v{c}}iak}  \& {Zhang}}{{Ursini} et~al.}{2022}]{2022MNRAS.510.3674U}
{Ursini} F.,  {Matt} G.,  {Bianchi} S.,  {Marinucci} A.,  {Dov{\v{c}}iak} M.,
  {Zhang} W.,  2022, \mn@doi [\mnras] {10.1093/mnras/stab3745}, \href
  {https://ui.adsabs.harvard.edu/abs/2022MNRAS.510.3674U} {510, 3674}

\bibitem[\protect\citeauthoryear{{Walter} \& {Courvoisier}}{{Walter} \&
  {Courvoisier}}{1990}]{1990A&A...233...40W}
{Walter} R.,  {Courvoisier} T.~J.~L.,  1990, \aap, \href
  {https://ui.adsabs.harvard.edu/abs/1990A&A...233...40W} {233, 40}

\bibitem[\protect\citeauthoryear{{Walter} \& {Courvoisier}}{{Walter} \&
  {Courvoisier}}{1992}]{1992A&A...258..255W}
{Walter} R.,  {Courvoisier} T.~J.~L.,  1992, \aap, \href
  {https://ui.adsabs.harvard.edu/abs/1992A&A...258..255W} {258, 255}

\bibitem[\protect\citeauthoryear{{Weisskopf}, {Elsner}  \&
  {O'Dell}}{{Weisskopf} et~al.}{2010}]{2010SPIE.7732E..0EW}
{Weisskopf} M.~C.,  {Elsner} R.~F.,   {O'Dell} S.~L.,  2010, in {Arnaud} M.,
  {Murray} S.~S.,   {Takahashi} T.,  eds,  Society of Photo-Optical
  Instrumentation Engineers (SPIE) Conference Series Vol. 7732, Space
  Telescopes and Instrumentation 2010: Ultraviolet to Gamma Ray. p. 77320E
  (\mn@eprint {arXiv} {1006.3711}), \mn@doi{10.1117/12.857357}

\bibitem[\protect\citeauthoryear{{Weisskopf} et~al.,}{{Weisskopf}
  et~al.}{2016}]{2016SPIE.9905E..17W}
{Weisskopf} M.~C.,  et~al., 2016, in {den Herder} J.-W.~A.,  {Takahashi} T.,
  {Bautz} M.,  eds,  Society of Photo-Optical Instrumentation Engineers (SPIE)
  Conference Series Vol. 9905, Space Telescopes and Instrumentation 2016:
  Ultraviolet to Gamma Ray. p. 990517, \mn@doi{10.1117/12.2235240}

\bibitem[\protect\citeauthoryear{{Whiteoak}}{{Whiteoak}}{1966}]{1966ZA.....64..181W}
{Whiteoak} J.~B.,  1966, \zap, \href
  {https://ui.adsabs.harvard.edu/abs/1966ZA.....64..181W} {64, 181}

\bibitem[\protect\citeauthoryear{{Yang}, {Yuan}, {Yuan}  \& {White}}{{Yang}
  et~al.}{2021}]{2021ApJ...914..131Y}
{Yang} H.,  {Yuan} F.,  {Yuan} Y.-F.,   {White} C.~J.,  2021, \mn@doi [\apj]
  {10.3847/1538-4357/abfe63}, \href
  {https://ui.adsabs.harvard.edu/abs/2021ApJ...914..131Y} {914, 131}

\bibitem[\protect\citeauthoryear{{Zhang} et~al.,}{{Zhang}
  et~al.}{2016}]{2016SPIE.9905E..1QZ}
{Zhang} S.~N.,  et~al., 2016, in {den Herder} J.-W.~A.,  {Takahashi} T.,
  {Bautz} M.,  eds,  Society of Photo-Optical Instrumentation Engineers (SPIE)
  Conference Series Vol. 9905, Space Telescopes and Instrumentation 2016:
  Ultraviolet to Gamma Ray. p. 99051Q (\mn@eprint {arXiv} {1607.08823}),
  \mn@doi{10.1117/12.2232034}

\bibitem[\protect\citeauthoryear{{Zhang} et~al.,}{{Zhang}
  et~al.}{2019a}]{2019SCPMA..6229502Z}
{Zhang} S.,  et~al., 2019a, \mn@doi [Science China Physics, Mechanics, and
  Astronomy] {10.1007/s11433-018-9309-2}, \href
  {https://ui.adsabs.harvard.edu/abs/2019SCPMA..6229502Z} {62, 29502}

\bibitem[\protect\citeauthoryear{{Zhang}, {Dov{\v{c}}iak}  \& {Bursa}}{{Zhang}
  et~al.}{2019b}]{2019ApJ...875..148Z}
{Zhang} W.,  {Dov{\v{c}}iak} M.,   {Bursa} M.,  2019b, \mn@doi [\apj]
  {10.3847/1538-4357/ab1261}, \href
  {https://ui.adsabs.harvard.edu/abs/2019ApJ...875..148Z} {875, 148}

\bibitem[\protect\citeauthoryear{{Zhang} et~al.,}{{Zhang}
  et~al.}{2019c}]{2019ApJ...876...49Z}
{Zhang} Z.-X.,  et~al., 2019c, \mn@doi [\apj] {10.3847/1538-4357/ab1099}, \href
  {https://ui.adsabs.harvard.edu/abs/2019ApJ...876...49Z} {876, 49}

\bibitem[\protect\citeauthoryear{{Zhang}, {Dov{\v{c}}iak}, {Bursa}, {Karas},
  {Matt}  \& {Ursini}}{{Zhang} et~al.}{2022}]{2022MNRAS.tmp.1852Z}
{Zhang} W.,  {Dov{\v{c}}iak} M.,  {Bursa} M.,  {Karas} V.,  {Matt} G.,
  {Ursini} F.,  2022, \mn@doi [\mnras] {10.1093/mnras/stac1937}, \href
  {https://ui.adsabs.harvard.edu/abs/2022MNRAS.tmp.1852Z} {515, 2882–2889}

\bibitem[\protect\citeauthoryear{{de Diego}, {Perez}, {Kidger}  \&
  {Takalo}}{{de Diego} et~al.}{1992}]{1992ApJ...396L..19D}
{de Diego} J.~A.,  {Perez} E.,  {Kidger} M.~R.,   {Takalo} L.~O.,  1992,
  \mn@doi [\apjl] {10.1086/186507}, \href
  {https://ui.adsabs.harvard.edu/abs/1992ApJ...396L..19D} {396, L19}

\makeatother
\end{thebibliography}








\bsp	
\label{lastpage}
\end{document}